\documentclass[journal,comsoc]{IEEEtran}
\usepackage[T1]{fontenc}
\usepackage{tikz}
\usepackage{graphicx}
\usepackage{amsthm}
\usepackage{braket}
\usepackage{cite}
\usepackage{optidef}
\newtheorem{lemma}{Lemma}

\usetikzlibrary{chains,arrows,fit,decorations}
\usepackage{setspace}

\usepackage{cite}

\ifCLASSINFOpdf
\else
\fi
\usepackage{amsmath}

\usepackage[cmintegrals]{newtxmath}
\usepackage{algorithm}
\usepackage{algorithmic}
\usepackage{epstopdf}
\begin{document}
	%
	\title{Bayesian User Localization and Tracking for Reconfigurable Intelligent Surface Aided MIMO Systems}
	\author{Boyu~Teng, Xiaojun~Yuan,~\IEEEmembership{Senior~Member,~IEEE},\\~Rui~Wang,~\IEEEmembership{Senior~Member,~IEEE},~and~Shi~Jin,~\IEEEmembership{Senior~Member,~IEEE}
	\thanks{B. Teng and X. Yuan are with the National Key Laboratory of Science and Technology on Communications, University of Electronic Science and Technology of China, Chengdu 610000, China (e-mail: byteng@std.uestc.edu.cn; xjyuan@uestc.edu.cn).}
	\thanks{R. Wang is with the College of Electronics and Information Engineering, Tongji University, Shanghai 201804, China. R. Wang is also with the Shanghai Institute of Intelligent Science and Technology, Tongji University, Shanghai 201804, China (e-mail: ruiwang@tongji.edu.cn).
}
	\thanks{S. Jin is with the National Mobile Communications Research Laboratory, Southeast University, Nanjing 210096, China, (e-mail:jinshi@seu.edu.cn).}
	\thanks{Corresponding author: Xiaojun Yuan.}}
 	\maketitle
	\begin{abstract}
	In this paper, we study the user localization and tracking problem in the reconfigurable intelligent surface (RIS) aided multiple-input multiple-output (MIMO) system, where a multi-antenna base station (BS) and multiple RISs are deployed to assist the localization and tracking of a multi-antenna user. By establishing a probability transition model for user mobility, we develop a message-passing algorithm, termed the Bayesian user localization and tracking (BULT) algorithm, to estimate and track the user position and the angle-of-arrival (AoAs) at the user in an online fashion. We also derive Bayesian Cram\'er Rao bound (BCRB) to characterize the fundamental performance limit of the considered tracking problem. To improve the tracking performance, we optimize the beamforming design at the BS and the RISs to minimize the derived BCRB. Simulation results show that our BULT algorithm can perform close to the derived BCRB, and significantly outperforms the counterpart algorithms without exploiting the temporal correlation of the user location.
	\end{abstract}
	

	%
	\IEEEpeerreviewmaketitle

	\section{Introduction}
	Integrated sensing and communication (ISAC) has been identified as a promising technology for the sixth-generation (6G) mobile communications \cite{wymeersch2021integration,zhang2021overview,liu2021integrated}, where the joint design of sensing and communication systems becomes necessary to meet the increasing demand for communication and sensing services \cite{yang2021integrated,liu2021survey}. High-precision localization, as a critical issue in sensing, shows great potential needs in the emerging 6G application scenarios such as internet of vehicles (IoV), augmented reality (AR), virtual reality (VR), unmanned aerial vehicle (UAV) communications, etc \cite{wymeersch2019radio,bourdoux20206g,saad2019vision}. With large antenna array and broad spectrum bandwidth, 6G communication signals enable high resolution in the angular and delay domains, which makes possible high-precision localization based on radio signals \cite{wymeersch2021integration,zhang2021overview,liu2021integrated,bourdoux20206g}.\par
	From the electromagnetic propagation theory, the diffraction effect of radio signals becomes weaker as the frequency increases \cite{wymeersch2021integration,zhang2021overview}. This phenomenon is particularly severe in 6G since 6G uses much higher frequency bands than earlier generations to support extremely high-speed communications. As a result, 6G communications rely heavily on the existence of line-of-sight (LoS) paths to maintain a sufficient receive power. As for localization, the position information of a target is mostly carried by the LoS path between the target and the radar, which is typically assumed to exist for reliable sensing \cite{liu2020two}. Unfortunately, in mobile communication scenarios, transmitters and receivers are usually blocked by obstacles due to highly complicated propagation environments. To address this issue in ISAC tasks, reconfigurable intelligent surface (RIS, a.k.a. intelligent reflecting surface) is introduced as an energy-efficient and cost-effective device for performance enhancement \cite{huang2019reconfigurable}. A RIS is composed of a large number of passive reflecting elements that independently adjust the incident signals by inducing controllable amplitude and additional phase changes, and can be deployed in the wireless environment to create a virtual-line-of-sight (VLoS) path. Moreover, with efficient passive beamforming design, a RIS is able to focus the reflected signals in a desired direction, so as to significantly enhance the communication quality via the VLoS link \cite{wu2021intelligent,liu2020matrix}. While the RIS aided communication system has been extensively studied (see, e.g., \cite{yan2020passive,pan2020intelligent,abeywickrama2020intelligent,ozdogan2019intelligent,yuan2021reconfigurable,wu2019beamforming,he2019cascaded} and the references therein), the research on RIS-aided sensing is still in its infancy stage. Early works include RIS aided high-precision user localization \cite{wymeersc2020beyond,elzanaty2021reconfigurable}, mobile target tracking \cite{zegrar2020general}, the beamforming design of RIS in sensing tasks \cite{liu2021reconfigurable}, etc.\par
	The user localization methods for RIS-aided multiple-input multiple-output (MIMO) systems have been recently studied in \cite{zhang2021metalocalization,lin2021channel,wang2021joint,he2020adaptive}. The authors in \cite{zhang2021metalocalization} proposed a RIS-aided indoor user positioning scheme based on the received signal strength. By iterative optimization of the RIS phase shifts, a high resolution signal strength map is established to locate the target. In \cite{lin2021channel}, a hierarchical-codebook based beam searching method was introduced to estimate the RIS-aided channel parameters which are further utilized to obtain accurate positioning. In \cite{wang2021joint}, the authors estimated the channel angle parameters by maximum likelihood and obtains the user position by combining the channel information of RISs. For the MIMO orthogonal frequency division multiplexing (OFDM) system, the authors in \cite{he2020adaptive} realised environment mapping and user localization by using the twin-RIS structure and exploiting array signal processing to obtain the channel parameters. In the above studies, user/target localization is realised in each time slot independently. This implies that the temporal correlation of the target location is not exploited in the design of the algorithms in \cite{zhang2021metalocalization,lin2021channel,wang2021joint,he2020adaptive}, which may significantly limit the localization performance.\par
	In this paper, we consider the user localization and tracking problem in a downlink MIMO system, where a multi-antenna base station (BS) and multiple RISs are deployed to assist the localization and tracking of a multi-antenna user. Specifically, the user receives the signals from the RISs and estimates the angle-of-arrivals (AoAs) based on the received signals. Each RIS acts as an anchor with its location known in advance. The user location is estimated by combining the angle information and the RIS location information. For user mobility, we build a probabilistic transition model of user movement by considering the fact that the user position changes continuously over time. Based on a factor-graph representation of the probability model, we develop a novel message-passing algorithm, termed the Bayesian user localization and tracking (BULT) algorithm, to efficiently solve the considered problem. In the algorithm design, various approximations are introduced to reduce the computational complexity involved in message passing.\par
    As a benchmark, we derive the Bayesian Cram\'er Rao bound (BCRB) that serves as a theoretic mean square error (MSE) lower bound of the considered estimation problem. We show by simulation results that our BULT algorithm performs close to the derived BCRB at relatively high signal-to-noise ratio (SNR), and that our algorithm significantly outperforms the counterpart algorithms without exploiting the temporal correlation of the user location.\par
    Furthermore, we consider the design of active beamforming at the BS and passive beamforming at the RISs. We develop an alternating algorithm for the design of active and passive beamforming to minimize the derived BCRB. Alternatively, to reduce computational complexity, we further propose a directional beamforming design at the BS and the RISs based on the estimated user location. We show that both beamforming strategies achieve similar performance, but the latter requires much lower complexity since it does not involve the iteration between active and passive beamforming optimizations.\par
    The remainder of this paper is organized as follows. In Section II, we introduce the RISs aided MIMO system and formulate the user localization and tracking problem. In Section III, we derive the message calculations and develop the Bayesian user localization and tracking algorithm. In Section IV, we analyse the BCRB for the tracking parameters. In Section V, we discuss the beamforming design for the BS and the RISs. Numerical results are presented in Section VI, and the paper concludes in Section VII.\par
	\textit{Notations:} Throughout, bold lowercase letters and bold capital letters are used to respectively denote vectors and matrices. We use $(\cdot)^{\mathrm{T}}$ and $(\cdot)^{\mathrm{H}}$ to denote the transpose and the conjugate transpose respectively. We use Tr($\mathbf{X}$) to denote the trace of $\mathbf{X}$, diag($\mathbf{x}$) to denote the diagonal matrix with its diagonal entries given by $\mathbf{x}$, $[\mathbf{X}]_{i,j}$ to denote the $(i,j)$-th term of $\mathbf{X}$, and $\mathbf{I}$ to denote the identity matrix. We use $\mathcal{N}\left(\mathbf{x};\boldsymbol{\mu},\boldsymbol{\Sigma}\right)$ and $\mathcal{CN}\left(\mathbf{x};\boldsymbol{\mu},\boldsymbol{\Sigma}\right)$ to denote the real Gaussian distribution and the circularly-symmetric Gaussian distribution with mean vector $\boldsymbol{\mu}$ and covariance matrix $\boldsymbol{\Sigma}$. We use $\mathbb{E}[\cdot]$ to denote the expectation operator, $\odot$ to denote Hadamard product, $\|\cdot\|_p$ to denote the $\ell_{p}$ norm. $\mathcal{R}\{\mathbf{X}\}$ and $\mathcal{I}\{\mathbf{X}\}$ are respectively the real part and the imaginary part of $\mathbf{X}$.
	\section{System Model and Problem Formulation}
	\label{Section2}
\subsection{System Description}
	We consider a multi-RIS aided MIMO system consisting of a BS, a user, and $K$ RISs, as illustrated in Fig. 1. The BS is equipped with $N_\mathrm{B}$ antennas, the user is equipped with $N_\mathrm{U}$ antennas, and each RIS consists of $N_\mathrm{R}$ reflecting elements. The antenna and reflecting elements are arranged as uniform linear arrays (ULAs) for the BS, the user, and the RISs; and the antenna/element interval is set to half of the carrier frequency wavelength $\lambda$. We assume that the VLoS path reflected by each RIS always exists while the LoS path propagated from the BS to the user is blocked by obstacles. We further assume downlink transmission where the transmitted signals of the BS are reflected by the RISs and then received by the user for localization and tracking. \par
	Assume that a 3D Cartesian coordinate system has been set up appropriately. The position of the BS, the $i$-th RIS and the user in time slot $t$ are denoted by the three-dimension vectors $\mathbf{p}_\mathrm{B}$, $\mathbf{p}_{\mathrm{R},i}$ and $\mathbf{p}_\mathrm{U}^{(t)}$ respectively. Further assume that the position of the BS and the RISs can be acquired by the user and the BS accurately in advance since the BS and the RISs are deployed stationarily.
	\begin{figure}
		\centering
		\resizebox{8cm}{!}{\includegraphics{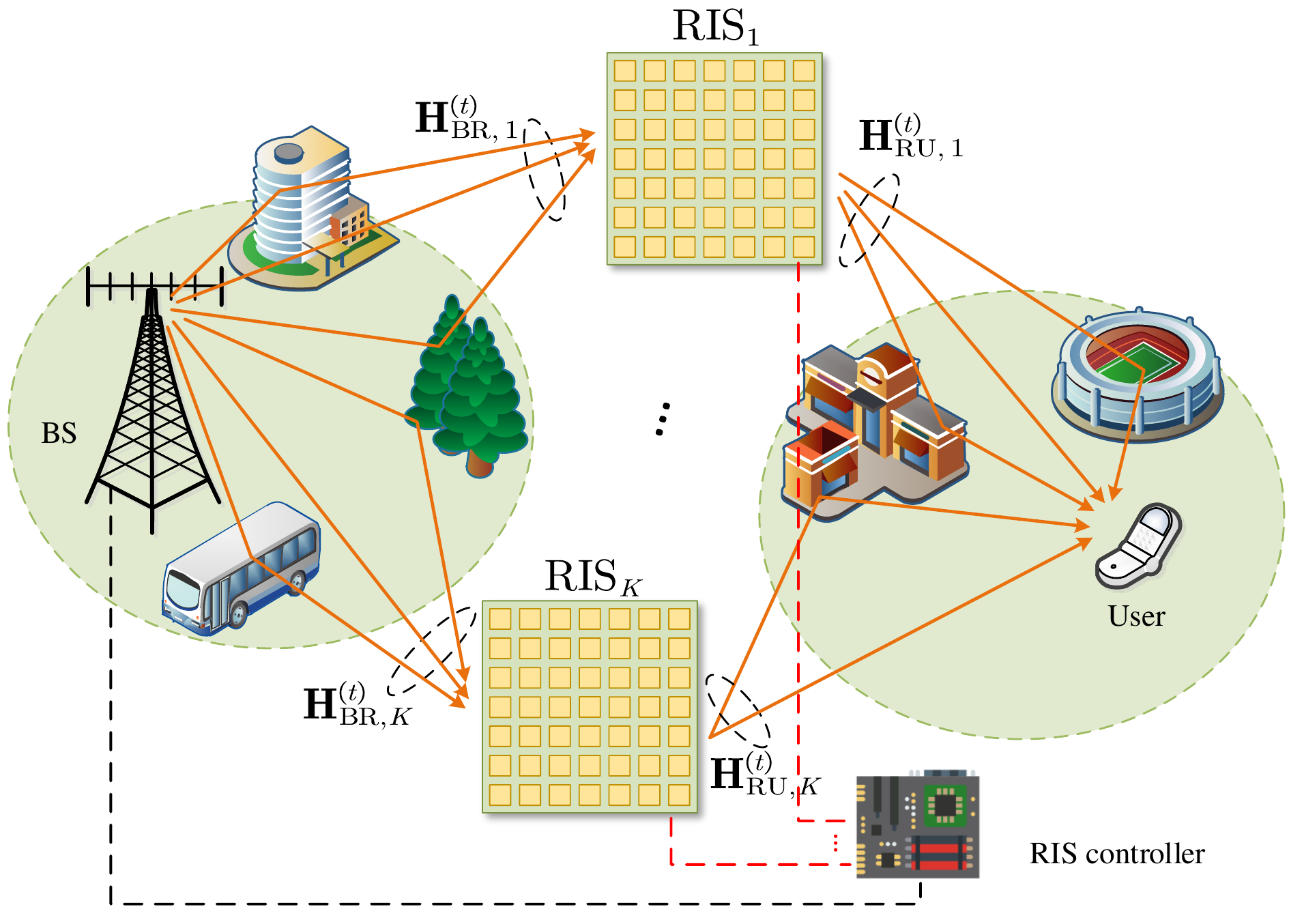}}
		\caption{System model of RISs aided MIMO system.}
		\label{fig_system}
	\end{figure}
\subsection{Channel Model}
    For the considered user tracking problem, the channel is in general correlated over time. The temporal correlation of the channel is presented in the next subsection by considering the geometric relationship of the user position and the channel parameters. In the following, we establish the geometric channel model for each time slot $t$. Specifically, the multi-path channel from the BS to the $i$-th RIS and from the $i$-th RIS to the user are denoted by $\mathbf{H}_{\mathrm{RB},i}^{(t)} \in \mathbb{C}^{N_\mathrm{R} \times N_\mathrm{B}}$ and $\mathbf{H}_{\mathrm{UR},i}^{(t)} \in \mathbb{C}^{N_\mathrm{U} \times N_\mathrm{R}}$ respectively, modeled as \cite{liu2020matrix}
	\begin{align}
	\label{racian_channel1}
	    \mathbf{H}_{\mathrm{RB},i}^{(t)} = \sqrt{\frac{\alpha}{1+\alpha}}\bar{\mathbf{H}}_{\mathrm{RB},i} +\sqrt{\frac{1}{1+\alpha}}\tilde{\mathbf{H}}_{\mathrm{RB},i}^{(t)},\\
	    \label{racian_channel2}
	    \mathbf{H}_{\mathrm{UR},i}^{(t)} = \sqrt{\frac{\alpha}{1+\alpha}}\bar{\mathbf{H}}_{\mathrm{UR},i}^{(t)} +\sqrt{\frac{1}{1+\alpha}}\tilde{\mathbf{H}}_{\mathrm{UR},i}^{(t)},
	\end{align}
	where $\tilde{\mathbf{H}}_{\mathrm{RB},i}^{(t)}\in \mathbb{C}^{N_\mathrm{R} \times N_\mathrm{B}}$ and $\tilde{\mathbf{H}}_{\mathrm{UR},i}^{(t)}\in \mathbb{C}^{N_\mathrm{U} \times N_\mathrm{R}}$ are the non-LoS components whose elements follow zero-mean complex Gaussian distributions; $\alpha$ is the Rician K-factor; $\bar{\mathbf{H}}_{\mathrm{RB},i}\in \mathbb{C}^{N_\mathrm{R} \times N_\mathrm{B}}$ and $\bar{\mathbf{H}}_{\mathrm{UR},i}^{(t)}\in \mathbb{C}^{N_\mathrm{U} \times N_\mathrm{R}}$ are the VLoS components given by
	\begin{align}
	    \label{LOS channel}
		\bar{\mathbf{H}}_{\mathrm{RB},i} =\rho _{\mathrm{B},i}  \mathbf{a}_{\mathrm{R}}( \theta _{\mathrm{R},i} ) \mathbf{a}_\mathrm{B}^{\mathrm{H}}(\vartheta _{\mathrm{B},i}), \\
		\bar{\mathbf{H}}_{\mathrm{UR},i}^{(t)} =\rho _{\mathrm{U},i}^{(t)}  \mathbf{a}_\mathrm{U}( \theta _{\mathrm{U},i}^{(t)} ) \mathbf{a}_\mathrm{R}^{\mathrm{H}}(\vartheta_{\mathrm{R},i}^{(t)}),
	\end{align}
	where $\rho _{\mathrm{B},i}$ and $\rho _{\mathrm{U},i}^{(t)}$ are the complex channel gains for the VLoS path; $ \theta _{\mathrm{R},i}$, $\vartheta _{\mathrm{B},i}$, $\theta_{\mathrm{U},i}^{(t)}$ and $\vartheta_{\mathrm{R},i}^{(t)}$ are respectively the cosine of the angle of arrival (AoA) at the $i$-th RIS, the cosine of the angle of departure (AoD) at the BS, the cosine of the AoA at the user and the cosine of the AoD at the $i$-th RIS, and $t$ is the time index;
	$\mathbf{a}_{\mathrm{R}}( \theta _{\mathrm{R},i} )$, $\mathbf{a}_{\mathrm{B}}(\vartheta _{\mathrm{B},i})$, $\mathbf{a}_{\mathrm{U}}( \theta _{\mathrm{U},i}^{(t)} )$ and $\mathbf{a}_{\mathrm{R}}(\vartheta _{\mathrm{R},i}^{(t)})$ are steering vectors in the form of line spectrum as
	\begin{equation}
		\mathbf{a}_{{x}}(\theta) = [1,e^{j\pi {\theta}},\dots,e^{j\pi (N_{x}-1) {\theta}}],\quad \mathrm{for}\; x\in\{\mathrm{B},\mathrm{R},\mathrm{U}\}.
	\end{equation}
We note that $\theta _{\mathrm{R},i}$ and $\vartheta _{\mathrm{B},i}$ are invariant over time and are assumed to be known beforehand based on the knowledge of $\mathbf{p}_\mathrm{B}$ and $\mathbf{p}_{\mathrm{R},i}$. From \eqref{racian_channel1} and \eqref{racian_channel2}, the received signal $\mathbf{y}^{(t)} \in \mathbb{C}^{N_\mathrm{U}}$ at the user antennas can be written as
	\begin{equation}
		\label{eq3}
		\mathbf{y}^{(t)} =\sum_{i=1}^K{\zeta_{i}^{(t)}\mathbf{H}_{\mathrm{UR},i}^{{(t)}} \mathbf{\Omega }_{i}^{{(t)}}\mathbf{H}_{\mathrm{RB},i}^{{(t)}}  \mathbf{f}^{(t)}}x(t)+\mathbf{w}^{{(t)}},
	\end{equation}
	where the pilot signal $x(t)$ is set by $x(t)=1$ for the user localization and tracking purpose; $\mathbf{f}^{{(t)}}\in \mathbb{C}^{N_\mathrm{B}}$ is the beamforming vector at BS side; $\mathbf{w}^{{(t)}} \sim \mathcal{CN}\left(\mathbf{w}^{{(t)}};\mathbf{0},\sigma_w^2\mathbf{I}\right)$ is the added Gaussian noise at user antennas; $\zeta_{i}^{(t)}$ is the reflection coefficient at RIS which generally varies in different AoAs and AoDs due to physical imperfection \cite{tang2020wireless}; $\mathbf{\Omega}_{i}^{{(t)}}=\mathrm{diag}(\boldsymbol{\omega}_{i}^{{(t)}})\in\mathbb{C}^{N_\mathrm{R}\times N_\mathrm{R}}$ is the diagonal phase shift matrix of the $i$-th RIS with the phase shift vector $\boldsymbol{\omega}_{i}^{{(t)}}=[{\omega}_{i,1}^{{(t)}},\dots,{\omega}_{i,N_{\mathrm{R}}}^{{(t)}}]$, where $|{\omega}_{i,n}^{{(t)}}|=1$ for $\forall i,n$. We assume that the phase shift of a reflecting element is continuously valued.\par
	In user tracking, we are interested in the channel parameters of the VLoS path, i.e., $\bar{\mathbf{H}}_{\mathrm{UR},i}$, with its key parameter $\theta_{\mathrm{U},i}^{{(t)}}$ related to the user position by the geometric constraint as
	\begin{equation}
		\label{eq10}
		\theta_{\mathrm{U},i}^{{(t)}}=\frac{\left(\mathbf{p}_{\mathrm{R},i}-\mathbf{p}_{\mathrm{U}}^{{(t)}}\right)^\mathrm{T}\mathbf{e}_\mathrm{U}}{\Vert \mathbf{p}_{\mathrm{R},i}-\mathbf{p}_{\mathrm{U}}^{{(t)}} \Vert_2},
	\end{equation}
	where $\mathbf{e}_\mathrm{U} $ is the unit direction vector of the user antennas and obtained by the built-in sensor in advance. Thus, we separate the VLoS component from the other signal components. Specifically, the received signal in \eqref{eq3} is further expressed as
	\begin{equation}
	    \mathbf{y}^{{(t)}} =\sum_{i=1}^K{\zeta_{i}^{(t)}\bar{\mathbf{H}}_{\mathrm{UR},i}^{{(t)}} \mathbf{\Omega }_{i}^{{(t)}}\bar{\mathbf{H}}_{\mathrm{RB},i}  \mathbf{f}^{(t)}}+\mathbf{n}^{{(t)}},
	\end{equation}
	where $\mathbf{n}^{{(t)}}\in\mathbb{C}^{N_\mathrm{U}}$ is the interference-plus-noise term given by
	\begin{align}
	    \mathbf{n}^{(t)}=&\sum_{i=1}^K{\zeta_{i}^{(t)}\left(\tilde{\mathbf{H}}_{\mathrm{RB},i}^{(t)} \mathbf{\Omega }_{i}^{(t)}\tilde{\mathbf{H}}_{\mathrm{RB},i}^{(t)}  \mathbf{f}^{(t)}+       \tilde{\mathbf{H}}_{\mathrm{UR},i}^{(t)} \mathbf{\Omega }_{i}^{(t)}\bar{\mathbf{H}}_{\mathrm{RB},i}  \mathbf{f}^{(t)}   \right.}&&\notag\\
	    &{\left.+\bar{\mathbf{H}}_{\mathrm{UR},i}^{{(t)}} \mathbf{\Omega }_{i}^{{(t)}}\tilde{\mathbf{H}}_{\mathrm{RB},i}^{{(t)}}\mathbf{f}^{(t)}\right)} + \mathbf{w}^{(t)}.
	\end{align}
	 Considering the severe path loss of the non-LoS path and the beamforming setting for the BS and the RISs provided in Section \ref{Section5}, we treat the non-LoS path signal as interference draw from a complex Gaussian distribution \cite{pan2020intelligent}, i.e., $\mathbf{n}^{{(t)}}\sim \mathcal{CN}\left(\mathbf{n}^{{(t)}};\mathbf{0},\sigma_n^2\mathbf{I}\right)$. Therefore, the received signal in time slot $t$ can be further simplified by
	 \begin{subequations}
	 	\begin{align}
	 		\label{eq8}
	 		\mathbf{y}^{(t)} \!\!=\! & \sum_{i=1}^K\!{\zeta_{i}^{(t)}\!\!\rho_{\mathrm{U},i}^{(t)}\rho_{\mathrm{B},i}{\mathbf{a}_{\mathrm{U}}\!( \theta _{\mathrm{U},i}^{{(t)}} )\mathbf{a}_{\mathrm{R}}^{\mathrm{H}}\!(\! \vartheta _{\mathrm{R},i}^{{(t)}}\! )}\mathbf{\Omega }_{i}^{(t)}\!{\mathbf{a}_{\mathrm{R}}\!(\! \theta _{\mathrm{R},i}\!)\mathbf{a}_{\mathrm{B}}^{\mathrm{H}}\!(\! \vartheta _{\mathrm{B},i}\! )}}\mathbf{f}^{(t)}\!\!+\!\mathbf{n}^{(t)}\\
	 		\label{eq9}
	 		= \!&\sum_{i=1}^K\!{\rho _{i}^{(t)}\mathbf{a}_{\mathrm{U}}(\theta _{\mathrm{U},i}^{{(t)}})}+\mathbf{n}^{(t)},
	 	\end{align}
	 \end{subequations}
where
	\begin{equation}
		\label{11}
		\rho_{i}^{(t)} = { \rho_{\mathrm{UB},i}^{(t)} \mathbf{a}_{\mathrm{r}}^{\mathrm{H}}( \vartheta _{\mathrm{R},i}^{{(t)}} )}\mathbf{\Omega }_{i}^{{(t)}}{\mathbf{a}_{\mathrm{R}}(\theta _{\mathrm{R},i}) \mathbf{a}_{\mathrm{B}}^{\mathrm{H}}( \vartheta _{\mathrm{B},i} )\mathbf{f}^{(t)}}
	\end{equation}
	is the equivalent complex path gain of the VLoS path for the $i$-th RIS with $\rho_{\mathrm{UB},i}^{(t)}=\zeta_{i}^{(t)}\rho_{\mathrm{U},i}^{{(t)}}\rho_{\mathrm{B},i}$. 
	We notice that the estimation of AoA in user $\theta _{\mathrm{U},i}^{{(t)}}$ falls into the category of line spectrum inference \cite{badiu2017variational}. The estimation of angle parameters can help with user tracking, based on which our user tracking method is developed.
	\subsection{Probabilistic Problem Formulation}
	\label{prob_form}
	In this section, we describe the user tracking problem by building a probability model for user position $\mathbf{p}_{\mathrm{U}}^{(t)}$, user AoAs $\{\theta _{\mathrm{U},i}^{{(t)}}\}$ and observation signal $\mathbf{y}^{(t)}$. 
	From \eqref{eq9}, the VLoS channel angle parameters $\theta _{\mathrm{U},i}^{(t)}$ corresponding to different RISs constitute the line spectrum of the received signal. With the geometric constraint in \eqref{eq10}, the conditional probability density function (pdf) $p(\theta_{\mathrm{U},i}^{{(t)}}|\mathbf{p}_\mathrm{U}^{(t)}) $ is represented as
	\begin{equation}
		\label{geometric_factornode}
		p(\theta_{\mathrm{U},i}^{{(t)}}|\mathbf{p}_\mathrm{U}^{(t)})=\delta\left(\theta _{\mathrm{U},i}^{{(t)}}-\frac{\left(\mathbf{p}_{\mathrm{R},i}-\mathbf{p}_\mathrm{U}^{(t)}\right)^\mathrm{T}\mathbf{e}_\mathrm{U}}{\Vert \mathbf{p}_{\mathrm{R},i}-\mathbf{p}_\mathrm{U}^{(t)} \Vert_2}\right),
	\end{equation}
	where $\delta(\cdot)$ is the Dirac delta function. Based on \eqref{eq9}, the likelihood function of $\boldsymbol{\theta}^{(t)}=[\theta _{\mathrm{U},1}^{{(t)}},\dots,\theta _{\mathrm{U},K}^{{(t)}}]^\mathrm{T}$ and $\boldsymbol{\rho}^{(t)}=[\rho_{1}^{{(t)}},\dots,\rho_{K}^{{(t)}}]^\mathrm{T}$ at time slot $t$ is given by
	\begin{equation}
		p(\mathbf{y}^{{(t)}}|\boldsymbol{\theta}^{(t)},\boldsymbol{\rho}^{(t)})=\mathcal{CN}\left(\mathbf{y}^{{(t)}};\sum_{i=1}^K{\rho_{i}^{{(t)}}\mathbf{a}_{\mathrm{U}}( \theta_{\mathrm{U},i}^{{(t)}} )},\sigma_n^2\mathbf{I}\right),
	\end{equation}
Due to the mobility of the user and the uncertainty of the reflection coefficient $\zeta_{i}^{(t)}$, we model the prior distribution of $\rho_{i}^{(t)}$ as a complex Gaussian distribution as
	\begin{equation}
		p(\rho_{i}^{(t)}) = \mathcal{CN}\left(\rho_{i}^{(t)};0,\sigma_\rho^2\right).
	\end{equation}
We assume that $\rho_{i}^{(t)}$ is independent of $\rho_{i'}^{(t')}$ for any $i \neq i'$ or $t \neq t'$.
	\par
	We now introduce a Markov probability transition model to help with the successive user tracking problem. Following \cite{va2016beam}\cite{zhang2016tracking}, we model the difference between the positions at any two adjacent time slots as an independent Gaussian noise, i.e.,
	\begin{equation}
		\mathbf{p}_\mathrm{U}^{(t)}=\mathbf{p}_\mathrm{U}^{(t-1)}+\mathbf{q}^{(t-1)},
	\end{equation}
where $\mathbf{q}^{(t-1)} \sim \mathcal{N}(0,\mathbf{C}_\mathrm{q})$ is the Gaussian transition noise. Thus, the conditional probability of $\mathbf{p}_\mathrm{U}^{(t)}$ given $\mathbf{p}_\mathrm{U}^{(t-1)}$ is given by
	\begin{equation}
		\label{eq11}
		p(\mathbf{p}_\mathrm{U}^{(t)}|\mathbf{p}_\mathrm{U}^{(t-1)})=\mathcal{N}(\mathbf{p}_\mathrm{U}^{(t)};\mathbf{p}_\mathrm{U}^{(t-1)},\mathbf{C}_\mathrm{q}).
	\end{equation}
	We assume that $\mathbf{q}^{(t)}$ and $\mathbf{q}^{(t')}$ are independent for any $t \neq t'$, i.e., the user positions $\{\mathbf{p}_\mathrm{U}^{(t)}\}$ form a Markov chain satisfying
	\begin{equation}
		p(\mathbf{p}_\mathrm{U}^{(t)}|\mathbf{p}_\mathrm{U}^{(1:t-1)},\mathbf{y}_\mathrm{U}^{(1:t-1)})=p(\mathbf{p}_\mathrm{U}^{(t)}|\mathbf{p}_\mathrm{U}^{(t-1)}),
	\end{equation}
where $\mathbf{p}_\mathrm{U}^{(t:t')}=\left[{(\mathbf{p}^{(t)}_\mathrm{U})}^\mathrm{T},...,{(\mathbf{p}_\mathrm{U}^{(t')})}^\mathrm{T}\right]^\mathrm{T}$ is the collection of user positions from time slot $t$ to time slot $t'$. Similar notation applies to $\mathbf{y}^{(t)}$, $\boldsymbol{\theta}^{(t)}$ and $\boldsymbol{\rho}^{(t)}$ for the received signals, the AoAs and the equivalent path gains.
	\par
	Based on the above discussions, the joint pdf of $\mathbf{p}_\mathrm{U}^{(0:t)}$, $\boldsymbol{\theta}^{(1:t)}$, $\boldsymbol{\rho}^{(1:t)}$ and $\mathbf{y}^{(1:t)}$ is given by
	\begin{align}
		\label{eq14}
		&p( \mathbf{p}_\mathrm{U}^{(0:t)},\boldsymbol{\theta}^{(1:t)}, \boldsymbol{\rho}^{(1:t)}, \mathbf{y}^{(1:t)} ) \notag\\
		&=p(\mathbf{p}_\mathrm{U}^{(0)}\!)\!\prod_{j=1}^t\!{p(\mathbf{p}_\mathrm{U}^{(j)}\!|\mathbf{p}_\mathrm{U}^{(j\!-\!1)}\!)p(\! \boldsymbol{\theta}^{(j)}\!|\mathbf{p}_\mathrm{U}^{(j)}\!)p( \!\boldsymbol{\rho}^{(j)}\! ) p(\mathbf{y}^{(j)}\!|\boldsymbol{\theta}^{(j)},\!\boldsymbol{\rho}^{(j)}\!)},
	\end{align}
	where $p(\!\boldsymbol{\theta}^{(j)}|\mathbf{p}_\mathrm{U}^{(j)})\!=\!\prod_{i=1}^{K}\!{p(\!{\theta}_{\mathrm{U},i}^{(j)}|\mathbf{p}_\mathrm{U}^{(j)})}$, $p(\boldsymbol{\rho}^{(j)})\!=\!\prod_{i=1}^{K}\!{p(\rho_{i}^{(j)})}$, and $p(\mathbf{p}_\mathrm{U}^{(0)})$ is the user position distribution at the initial time slot. Then, our localization and tracking problem is to estimate $\mathbf{p}_\mathrm{U}^{(t)}$ in each time slot $t$ given the historical received signal $\mathbf{y}^{(1:t)}$ and the initial position $\mathbf{p}_\mathrm{U}^{(0)}$.
	With the joint pdf given in \eqref{eq14}, following the Bayes' theorem, the posterior distribution of user position $\mathbf{p}_\mathrm{U}^{(t)}$ is given by
	\begin{align}
		\label{17_n}
	    &p(\mathbf{p}_\mathrm{U}^{(t)}|\mathbf{y}^{(1:t)},\mathbf{p}_\mathrm{U}^{(0)}) \notag\\
	    &\quad=\int{\frac{p(\mathbf{p}_\mathrm{U}^{(0:t)}, \boldsymbol{\theta}^{(1:t)}, \boldsymbol{\rho}^{(1:t)}, \mathbf{y}^{(1:t)} )}{p(\mathbf{y}^{(1:t)})p(\mathbf{p}_\mathrm{U}^{(0)})}\mathrm{d}\boldsymbol{\theta}^{(1:t)}\mathrm{d}\boldsymbol{\rho}^{(1:t)}\mathrm{d}\mathbf{p}_\mathrm{U}^{(1:t-1)}}
	\end{align}
Then, an online estimate of $\mathbf{p}_\mathrm{U}^{(t)}$ can be obtained by following the minimum mean-square error (MMSE) or maximum \textit{a posteriori} (MAP) principle. However, exact posterior estimation is generally intractable due to prohibitively high computational complexity caused by the integral in \eqref{17_n}. Thus, we resort to a low-complexity solution by following the message-passing principle, as detailed in the subsequent section.
	

	\section{Bayesian User Localization and Tracking Algorithm}
\label{Section3}
	\subsection{Factor Graph Representation}
\begin{figure}
	\centering
	\resizebox{\columnwidth}{!}{\begin{tikzpicture}
	\node (r1) at (4.4,2.3){} ;
	\draw (0,0) node[left=0.2] {$\mathbf{p}_\mathrm{U}^{(t)}$} circle [radius=0.3];
	\draw (0,3.6) node[left=0.2] {$\mathbf{p}_\mathrm{U}^{(t+1)}$} circle [radius=0.3];
	\draw (0,-3.6) node[left=0.2] {$\mathbf{p}_\mathrm{U}^{(t-1)}$} circle [radius=0.3];	
	\draw[fill=black] (-0.25,1.55) rectangle node[left=0.2cm]  (l1) {$p(\mathbf{p}_\mathrm{U}^{(t+1)}|\mathbf{p}_\mathrm{U}^{(t)})$} (0.25,2.05);
	\draw[fill=black] (-0.25,-2.05) rectangle node[left=0.2cm] {$p(\mathbf{p}_\mathrm{U}^{(t)}|\mathbf{p}_\mathrm{U}^{(t-1)})$} (0.25,-1.55);
	
	\draw (0,4.6) node (l2) {\rotatebox{90}{$\dots$}};
	\draw (0,-4.5) node (l3) {\rotatebox{90}{$\dots$}};
	\draw (-4,0) node {\rotatebox{90}{\text{Forward tracking}}};
	
	\draw[fill=black] (2.1,0.4) rectangle node[below=0.2cm] {$p(\theta _{\mathrm{U},2}^{(t)}|\mathbf{p}_\mathrm{U}^{(t)})$} (2.6,0.9);
	\draw[fill=black] (2.1,-0.9) rectangle node[below=0.2cm]  {$p(\theta _{\mathrm{U},3}^{(t)}|\mathbf{p}_\mathrm{U}^{(t)})$} (2.6,-0.4);
	\draw[fill=black] (2.1,1.75) rectangle node[below=0.2cm]  {$p(\theta _{\mathrm{U},1}^{(t)}|\mathbf{p}_\mathrm{U}^{(t)})$} (2.6,2.25);
	\draw[fill=black] (2.1,-2.25) rectangle node[below=0.2cm] (l4) {$p(\theta _{\mathrm{U},4}^{(t)}|\mathbf{p}_\mathrm{U}^{(t)})$} (2.6,-1.75);
	
	\draw (4.7,0.65) node[below = 0.2cm] {$\theta _{\mathrm{U},2}^{(t)}$} circle [radius=0.3];
	\draw (4.7,2) node[below = 0.2cm]  {$\theta _{\mathrm{U},1}^{(t)}$} circle  [radius=0.3];
	\draw (4.7,-0.65) node[below = 0.2cm] {$\theta _{\mathrm{U},3}^{(t)}$} circle [radius=0.3];
	\draw (4.7,-2) node[below = 0.2cm] (r2) {$\theta _{\mathrm{U},4}^{(t)}$} circle [radius=0.3];
	
	\draw[fill=black] (6.8,-0.25) rectangle node[right=0.2cm] (r3) {$p(\mathbf{y}^{(t)}|\boldsymbol{\theta}^{(t)},\boldsymbol{\rho}^{(t)})$} (7.3,0.25);
	
	\draw (7.05,-1) node[right = 0.2cm]  {$\boldsymbol{\rho}^{(t)}$} circle [radius=0.3];
	\draw[fill=black] (6.8,-2.25) rectangle node[right=0.2cm]   {$p(\boldsymbol{\rho}^{(t)})$} (7.3,-1.75);

	\draw (0,0.3) -- (0,1.55);
	\draw (0,-0.3) -- (0,-1.55);
	
	\draw (0,2.05) -- (0,3.3);
	\draw (0,-2.05) -- (0,-3.3);
	
	\draw (0,3.9) -- (0,4.2);
	\draw (0,-3.9) -- (0,-4.2);
	
	\draw (0.3,0) -- (2.1,0.65);
	\draw (0.3,0) -- (2.1,-0.65);
	\draw (0.3,0) -- (2.1,2);
	\draw (0.3,0) -- (2.1,-2);
	
	\draw (2.6,0.65) -- (4.4,0.65);
	\draw (2.6,-0.65) -- (4.4,-0.65);
	\draw (2.6,2) -- (4.4,2);
	\draw (2.6,-2) -- (4.4,-2);
	
	\draw (5.0,0.65) -- (6.8,0);
	\draw (5.0,-0.65) -- (6.8,0);
	\draw (5.0,2) -- (6.8,0);
	\draw (5.0,-2) -- (6.8,0);
	
	\draw (7.05,-0.25) -- (7.05,-0.7);
	\draw (7.05,-1.3) -- (7.05,-1.75);
	
	\draw (0.3,3.6) -- (1.5,3.75);
	\draw (0.3,3.6) -- (1.5,4.1);
	\draw (0.3,3.6) -- (1.5,3.45);
	\draw (0.3,3.6) -- (1.5,3.15);
	
	\draw (0.3,-3.6) -- (1.5,-3.75);
	\draw (0.3,-3.6) -- (1.5,-4.1);
	\draw (0.3,-3.6) -- (1.5,-3.45);
	\draw (0.3,-3.6) -- (1.5,-3.15);
	
	\draw[-latex,very thick] (-3.5,-3) -- (-3.5,3);
	
	\tikzset{blue dotted/.style={draw=black!50!white, line width=1pt,
			dash pattern=on 1pt off 4pt on 6pt off 4pt,
			inner sep=2mm, rectangle, rounded corners}};
	\node (first dotted box) [blue dotted, fit = (l1) (l2) (l3) (l4)] {};
	\node (second dotted box) [blue dotted, fit = (r1) (r2) (r3)] {};
	\node at (first dotted box.north) [above, inner sep=2mm] {\textbf{Position Tracking Module}};
	\node at (second dotted box.north) [above, inner sep=2mm] {\textbf{AoA estimation module}};
\end{tikzpicture}}
	
	\caption{Factor graph representation of \eqref{eq14} for an example in the time slot $t$ and $K=4$. This factor graph is divided into two modules: the position tracking module and the AoA estimation module.}
	\label{fig_factorgraph}
	\end{figure}
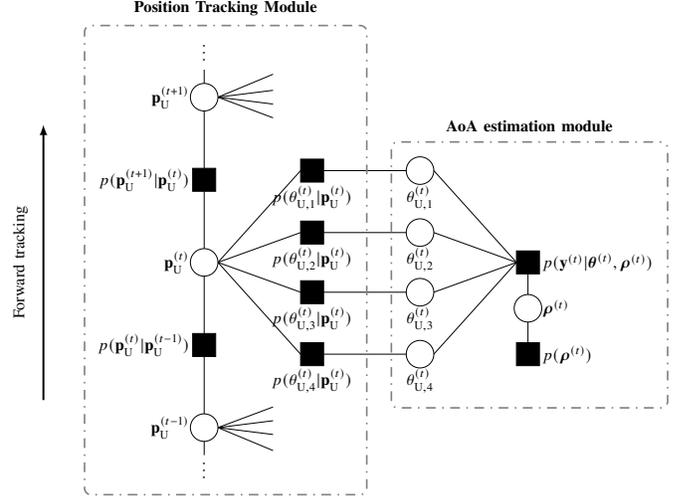

	In this section, we introduce our BULT algorithm based on the message passing principle. The factor graph corresponding to \eqref{eq14} is constructed as shown in Fig. \ref{fig_factorgraph}, where each blank circle represents a variable node, and each black rectangle represents a factor node. A variable node is connected to a factor node if the factor contains the variable. The variable node $\mathbf{y}^{(t)}$ is omitted since $\mathbf{y}^{(t)}$ is observed. In the factor graph, the position tracking module and AoA estimation module are introduced to illustrate the BULT algorithm. Specifically, the position tracking model includes the probability factors for the Markov chain of $\{\mathbf{p}_{\mathrm{U}}^{(t)}\}$; the AoA estimation module includes the probability factors for the receive signal $\mathbf{y}^{(t)}$ and the user AoAs $\{\theta_{\mathrm{U},i}^{(t)}\}$. These two modules are connected by factor node $p(\theta_{\mathrm{U},i}^{(t)}|\mathbf{p}_\mathrm{U}^{(t)})$ and variable node $\theta_{\mathrm{U},i}^{(t)}$ according to the geometric constraint. For notational brevity, we represent the factor node $p(\theta_{\mathrm{U},i}^{(t)}|\mathbf{p}_\mathrm{U}^{(t)})$ by $\varphi_{i}^{(t)}$ and the factor node  $p(\mathbf{p}_\mathrm{U}^{(t)}|\mathbf{p}_\mathrm{U}^{(t-1)})$ by $\psi^{(t)}$ in the subscript of a message. Denote by $\Delta_{a \rightarrow b}(\cdot)$ the message from node $a$ to $b$, and by $\mathbf{m}_{{a \rightarrow b}}$ and $\mathbf{C}_{{a \rightarrow b}}$ the mean vector and the covariance matrix of message $\Delta_{a \rightarrow b}(\cdot)$, respectively. Denote by $\Delta_{a}(\cdot)$ the message of variable node $a$.
\subsection{AoA Estimation Module}
The AoA estimation module is designed to provide the estimations of user AoAs $\{\theta_{\mathrm{U},i}^{(t)}\}$ which are further utilized in the position tracking module. The messages in the AoA estimation module are described below.
\subsubsection{Messages from $\theta_{\mathrm{U},i}^{(t)}$ to $\varphi_{i}^{(t)}$} Before the further discussion on the message calculation, we first give the pdf of a Von Mises (VM) distribution $\mathcal{M}(\theta;\mu,\kappa)$ as
\begin{equation}
	\label{VM}
	\mathcal{M}(\theta;\mu,\kappa)=\frac{1}{2\pi I_0 (\kappa)}\exp(\kappa \cos(\theta-\mu)),
\end{equation}
where $I_0$ denotes the modified Bessel function of the first kind in order $0$, $\mu$ and $\kappa$ are the mean direction and concentration parameters respectively.\par
For $\forall t$, $1 \le i\le K$, following the sum-product rule, the message from variable node $\theta_{\mathrm{U},i}^{(t)}$ to factor node $\varphi_{i}^{(t)}$ is given by
\begin{align}
	&\Delta_{\theta_{\mathrm{U},i}^{(t)}\rightarrow\varphi_{i}^{(t)}}(\theta_{\mathrm{U},i}^{(t)})\notag\\
	&\quad\propto\int_{\boldsymbol{\theta}^{(t)}_{\backslash i}}\int_{\boldsymbol{\rho}^{(t)}}{p(\mathbf{y}^{{(t)}}|\boldsymbol{\theta}^{(t)},\boldsymbol{\rho}^{(t)})p(\boldsymbol{\rho}^{(t)})\prod_{j\neq i}{\Delta _{\varphi_{i}^{(t)}\rightarrow \theta_{\mathrm{U},j}^{(t)}}(\theta_{\mathrm{U},j}^{(t)})}},
\end{align}
where $\boldsymbol{\theta}^{(t)}_{\backslash i}$ denotes the set includes all the entries in $\boldsymbol{\theta}^{(t)}$ except the $i$-th one, the message $\Delta _{\varphi_{i}^{(t)}\rightarrow \theta_{\mathrm{U},j}^{(t)}}(\theta_{\mathrm{U},j}^{(t)})$ can be treated as a prior knowledge of $\theta_{\mathrm{U},j}^{(t)}$ and is approximated by a VM distribution as discussed in next subsection. We further express $\Delta_{\theta_{\mathrm{U},i}^{(t)}\rightarrow\varphi_{i}^{(t)}}(\theta_{\mathrm{U},i}^{(t)})$ as
\begin{subequations}
	\label{22_n}
	\begin{align}
		&\Delta_{\theta_{\mathrm{U},i}^{(t)}\rightarrow\varphi_{i}^{(t)}}(\theta_{\mathrm{U},i}^{(t)})&&\notag\\
		&\quad\propto\frac{\int_{\boldsymbol{\theta}^{(t)}_{\backslash i}}\!\int_{\boldsymbol{\rho}^{(t)}}\!{p(\mathbf{y}^{{(t)}}|\boldsymbol{\theta}^{(t)}\!,\!\boldsymbol{\rho}^{(t)})p(\boldsymbol{\rho}^{(t)})\!\prod_{j}\!{\Delta _{\varphi_{i}^{(t)}\!\rightarrow\! \theta_{\mathrm{U},j}^{(t)}}(\theta_{\mathrm{U},j}^{(t)})}}}{\Delta _{\varphi_{i}^{(t)}\rightarrow \theta_{\mathrm{U},i}^{(t)}}(\theta_{\mathrm{U},i}^{(t)})}\label{33_n}\\
		&\quad\propto\frac{p(\theta_{\mathrm{U},i}^{(t)}|\mathbf{y}^{(t)})}{\Delta _{\varphi_{i}^{(t)}\rightarrow \theta_{\mathrm{U},i}^{(t)}}(\theta_{\mathrm{U},i}^{(t)})},\label{22b}
	\end{align}
\end{subequations}
where the integral in \eqref{33_n} is approximated by the posterior estimation of $\theta_{\mathrm{U},i}^{(t)}$ as $p(\theta_{\mathrm{U},i}^{(t)}|\mathbf{y}^{(t)})$. We resort to the variational Bayesian method to obtain $p(\theta_{\mathrm{U},i}^{(t)}|\mathbf{y}^{(t)})$.
By modeling the pdf of each AoA as a VM distribution, the VALSE algorithm is developed in \cite{badiu2017variational} by following the variational inference principle to obtain the approximated posterior estimates of the AoAs $\{\theta_{\mathrm{U},i}^{(t)}\}$ and the equivalent path gains $\{\rho^{(t)}_i\}$. Also, the model parameters $\sigma_\rho^{2}$ and $\sigma_n^{2}$ are iteratively estimated by the expectation maximization method.
We denote the mentioned VM distributions as 
\begin{subequations}
	\label{23}
	\begin{align}
		p(\theta_{\mathrm{U},i}^{(t)}|\mathbf{y}^{(t)}) &=  \mathcal{M}\left(\pi\theta_{\mathrm{U},i}^{{(t)}};\mu_{\theta_{\mathrm{U},i}^{(t)}},\kappa_{\theta_{\mathrm{U},i}^{(t)}}\right),\\
		\Delta _{\varphi_{i}^{(t)}\rightarrow \theta_{\mathrm{U},i}^{(t)}}(\theta_{\mathrm{U},i}^{(t)})&=\mathcal{M}\left(\pi\theta_{\mathrm{U},i}^{{(t)}};\mu_{\varphi_{i}^{(t)}\rightarrow \theta_{\mathrm{U},i}^{(t)}},\kappa_{\varphi_{i}^{(t)}\rightarrow \theta_{\mathrm{U},i}^{(t)}}\right),
	\end{align}
\end{subequations}
where the variable $\theta_{\mathrm{U},i}^{{(t)}}$ is multiplied by a constant $\pi$ to obey the VM distribution. From \cite{mardia2009directional}, the family of the VM distribution is closed under multiplication. Then, the message $\Delta_{\theta_{\mathrm{U},i}^{(t)}\rightarrow\varphi_{i}^{(t)}}(\theta_{\mathrm{U},i}^{(t)})$ is further approximated by
\begin{equation}
	\label{36}
	\Delta _{\theta_{\mathrm{U},i}^{(t)}\rightarrow \varphi_{i}^{(t)}}(\theta_{\mathrm{U},i}^{(t)}) \propto \mathcal{M}\left(\pi\theta_{\mathrm{U},i}^{{(t)}};\mu _{\theta_{\mathrm{U},i}^{(t)}\rightarrow \varphi_{i}^{(t)}},\kappa _{\theta_{\mathrm{U},i}^{(t)}\rightarrow \varphi_{i}^{(t)}} \right),
\end{equation}
where $\mu_{\theta_{\mathrm{U},i}^{(t)}\rightarrow \varphi_{i}^{(t)}}$ and $\kappa_{\theta_{\mathrm{U},i}^{(t)}\rightarrow \varphi_{i}^{(t)}}$ have a relationship of
\begin{align}
	\label{37}
	&\kappa_{\theta_{\mathrm{U},i}^{(t)}\rightarrow \varphi_{i}^{(t)}}\exp{\left(j\mu_{\theta_{\mathrm{U},i}^{(t)}\rightarrow \varphi_{i}^{(t)}}\right)} \notag\\
	&\quad=\kappa_{\theta_{\mathrm{U},i}^{(t)}}\exp{\left(j\mu_{\theta_{\mathrm{U},i}^{(t)}}\right)} - \kappa_{\varphi_{i}^{(t)}\rightarrow \theta_{\mathrm{U},i}^{(t)}}\exp{\left(j\mu_{\varphi_{i}^{(t)}\rightarrow \theta_{\mathrm{U},i}^{(t)}}\right)}.
\end{align}
The calculation of message $\Delta _{\theta_{\mathrm{U},i}^{(t)}\rightarrow \varphi_{i}^{(t)}}$ in \eqref{37} can be regarded as the extrinsic message calculation of the AoA estimation module; see more discussions on extrinsic messages, e.g, in \cite{xue2021denoising}.

\subsection{Position Tracking Module}
The position tracking module aims to estimate the user position $\mathbf{p}_{\mathrm{U}}^{(t)}$ based on the estimated AoAs. The messages in the position tracking module are described below.
\subsubsection{Messages from $\varphi_{i}^{(t)}$ to $\mathbf{p}_{\mathrm{U}}^{(t)}$}For $\forall t$, $1 \le i\le K$, the message from factor node $\varphi_{i}^{(t)}$ to variable node $\mathbf{p}_{\mathrm{U}}^{(t)}$ is given by
\begin{equation}
	\label{36_n}
	\Delta _{\varphi_{i}^{(t)}\rightarrow \mathbf{p}_{\mathrm{U}}^{(t)}}(\mathbf{p}_{\mathrm{U}}^{(t)}) \propto \int{p(\theta_{\mathrm{U},i}^{(t)}|\mathbf{p}_{\mathrm{U}}^{(t)})\Delta_{\theta_{\mathrm{U},i}^{(t)} \rightarrow\varphi_{i}^{(t)} }(\theta_{\mathrm{U},i}^{(t)})}, 
\end{equation}
where $\Delta_{\theta_{\mathrm{U},i}^{(t)} \rightarrow\varphi_{i}^{(t)} }(\theta_{\mathrm{U},i}^{(t)})$ is approximated as the VM distribution given in \eqref{36}. Therefore, \eqref{36_n} can be expressed as
\begin{align}
	\label{n27}
	&\Delta_{\varphi_{i}^{(t)}\rightarrow \mathbf{p}_{\mathrm{U}}^{(t)}}(\mathbf{p}_{\mathrm{U}}^{(t)}) \notag\\
	&\quad\propto \exp \left( \kappa _{\theta_{\mathrm{U},i}^{(t)}\rightarrow \varphi_{i}^{(t)}}\cos\left(\frac{\pi \left(\mathbf{p}_{\mathrm{R},i}-\mathbf{p}_{\mathrm{U}}^{(t)}\right)^{\mathrm{T}}\mathbf{e}_\mathrm{U}}{\left\| \mathbf{p}_{\mathrm{R},i}-\mathbf{p}_{\mathrm{U}}^{(t)} \right\|}-\mu_{\theta_{\mathrm{U},i}^{(t)}\rightarrow \varphi_{i}^{(t)}}\right) \right).
\end{align}
\subsubsection{Messages along the Markov chain $\{\mathbf{p}_\mathrm{U}^{(t)}\}$}
For $\forall t$, the message from variable node $\mathbf{p}_{\mathrm{U}}^{(t)}$ to factor node $\psi^{(t+1)}$ is given by
\begin{equation}
	\label{22}
	\Delta _{\mathbf{p}_{\mathrm{U}}^{(t)}\rightarrow\psi^{(t+1)} }(\mathbf{p}_{\mathrm{U}}^{(t)})\propto\Delta _{\psi^{(t)}\rightarrow\mathbf{p}_{\mathrm{U}}^{(t)} }(\mathbf{p}_{\mathrm{U}}^{(t)}) \mathcal{G}^{(t)}(\mathbf{p}_{\mathrm{U}}^{(t)}),
\end{equation}
where
\begin{equation}
	\label{30}
	\mathcal{G}^{(t)}(\mathbf{p}_{\mathrm{U}}^{(t)}) = \prod_{j=1}^K{\Delta_{\varphi_{j}^{(t)} \rightarrow \mathbf{p}_{\mathrm{U}}^{(t)}}(\mathbf{p}_{\mathrm{U}}^{(t)})}.
\end{equation}
Based on the central limit theorem \cite{donoho2009message}, the message $\mathcal{G}^{(t)}(\mathbf{p}_{\mathrm{U}}^{(t)})$ can be approximated as a Gaussian distribution as
\begin{equation}
	\label{30_n}
	\mathcal{G}^{(t)}(\mathbf{p}_{\mathrm{U}}^{(t)})=\mathcal{N}(\mathbf{p}_{\mathrm{U}}^{(t)};\mathbf{m}_{\mathcal{G}^{(t)}},\mathbf{C}_{\mathcal{G}^{(t)}}),
\end{equation}
where the mean vector $\mathbf{m}_{\mathcal{G}^{(t)}}$ and the covariance matrix $\mathbf{C}_{\mathcal{G}^{(t)}}$ are obtained by using the gradient descent method (GDM) and the Taylor series expansion. The detailed derivation can be found in Appendix \ref{taylor_series}.\par
For $\forall t$, the message from factor node $\psi^{(t)}$ to variable node $\mathbf{p}_{\mathrm{U}}^{(t)}$ is given by
\begin{equation}
	\label{21}
	\Delta _{\psi^{(t)}\rightarrow\mathbf{p}_{\mathrm{U}}^{(t)} }(\mathbf{p}_\mathrm{U}^{(t)})\propto\int_{\mathbf{p}_{\mathrm{U}}^{(t-1)}}{\Delta _{\mathbf{p}_{\mathrm{U}}^{(t-1)}\rightarrow\psi^{(t)} }(\mathbf{p}_{\mathrm{U}}^{(t-1)}) p(\mathbf{p}_{\mathrm{U}}^{(t)}|\mathbf{p}_{\mathrm{U}}^{(t-1)})}.
\end{equation}
We note that the factor nodes $\{\psi^{(t)}\}$ are in the form of Gaussian distribution as given in \eqref{eq11}. By assuming that a Gaussian message is provided in the initial time slot, all the messages passing along the Markov chain are Gaussian since the product of two Gaussian messages is still Gaussian. Therefore, the mean vector and covariance matrix of message $\Delta _{\mathbf{p}_{\mathrm{U}}^{(t)}\rightarrow\psi^{(t+1)} }(\mathbf{p}_{\mathrm{U}}^{(t)})$ are respectively given by
\begin{subequations}
	\label{32_nn}
	\begin{align}
		\mathbf{m}_{\mathbf{p}_{\mathrm{U}}^{(t)}\!\rightarrow\psi^{(t+1)}}&=\mathbf{C}_{\mathbf{p}_{\mathrm{U}}^{(t)}\!\rightarrow\psi^{(t+1)}}\!\left(\!\mathbf{C}_{\mathcal{G}^{(t)}}^{-1}\mathbf{m}_{\mathcal{G}^{(t)}} \!+\!\mathbf{C}_{\psi^{(t)}\!\rightarrow\mathbf{p}_{\mathrm{U}}^{(t)}}^{-1}\mathbf{m}_{\psi^{(t)}\!\rightarrow\mathbf{p}_{\mathrm{U}}^{(t)}}\!\right),\\
		 \mathbf{C}_{\mathbf{p}_{\mathrm{U}}^{(t)}\!\rightarrow\psi^{(t+1)}}^{-1}&= \mathbf{C}_{\mathcal{G}^{(t)}}^{-1} + \mathbf{C}_{\psi^{(t)}\rightarrow\mathbf{p}_{\mathrm{U}}^{(t)}}^{-1},
	\end{align}
\end{subequations}
where according to \eqref{eq11} and \eqref{21} we have
\begin{subequations}
	\label{33_nn}
	\begin{align}
		\mathbf{m}_{\psi^{(t)}\rightarrow\mathbf{p}_{\mathrm{U}}^{(t)}} &= \mathbf{m}_{\mathbf{p}_{\mathrm{U}}^{(t-1)}\rightarrow\psi^{(t)}},\\
		\mathbf{C}_{\psi^{(t)}\rightarrow\mathbf{p}_{\mathrm{U}}^{(t)}} &= \mathbf{C}_{\mathbf{p}_{\mathrm{U}}^{(t-1)}\rightarrow\psi^{(t)}} + \mathbf{C}_{\mathrm{q}}.
	\end{align}
\end{subequations} 
Based on \eqref{32_nn} and \eqref{33_nn}, the mean vectors and covariance matrices of the Gaussian messages along the Markov chain can be recursively calculated.
\subsubsection{Messages from $\mathbf{p}_{\mathrm{U}}^{(t)}$ to $\varphi_{i}^{(t)}$}For $\forall t$, $1 \le i\le K$, the message from variable node $\mathbf{p}_{\mathrm{U}}^{(t)}$ to factor node $\varphi_{i}^{(t)}$ is given by
\begin{equation}
	\Delta _{\mathbf{p}_{\mathrm{U}}^{(t)}\rightarrow\varphi_{i}^{(t)} }(\mathbf{p}_{\mathrm{U}}^{(t)})\propto \Delta _{\psi^{(t)}\rightarrow\mathbf{p}_{\mathrm{U}}^{(t)} }(\mathbf{p}_{\mathrm{U}}^{(t)})\mathcal{G}_{\backslash i}^{(t)}(\mathbf{p}_{\mathrm{U}}^{(t)}),
\end{equation}
where
\begin{equation}
	\label{36_nn}
	\mathcal{G}_{\backslash i}^{(t)}(\mathbf{p}_{\mathrm{U}}^{(t)})= \prod_{j\ne i}{\Delta_{\varphi_{j}^{(t)} \rightarrow \mathbf{p}_{\mathrm{U}}^{(t)}}(\mathbf{p}_{\mathrm{U}}^{(t)})}
\end{equation}
differs from \eqref{30} for only one term and can be approximated as a Gaussian distribution with mean vector $\mathbf{m}_{\mathcal{G}_{\backslash i}^{(t)}}$ and the covariance matrix $\mathbf{C}_{\mathcal{G}_{\backslash i}^{(t)}}$ similarly. Therefore,
$\Delta _{\mathbf{p}_{\mathrm{U}}^{(t)}\rightarrow\varphi_{i}^{(t)} }(\mathbf{p}_{\mathrm{U}}^{(t)})$ is also a Gaussian message with its mean vector and covariance matrix calculated by
\begin{subequations}
	\label{29}
	\begin{align}
		\mathbf{m}_{\mathbf{p}_{\mathrm{U}}^{(t)}\rightarrow\varphi_{i}^{(t)}}&=\mathbf{C}_{\mathbf{p}_{\mathrm{U}}^{(t)}\rightarrow\varphi_{i}^{(t)}}\left(\mathbf{C}_{\mathcal{G}_{\backslash i}^{(t)}}^{-1}\mathbf{m}_{\mathcal{G}_{\backslash i}^{(t)}} +\mathbf{C}_{\psi^{(t)}\rightarrow\mathbf{p}_{\mathrm{U}}^{(t)}}^{-1}\mathbf{m}_{\psi^{(t)}\rightarrow\mathbf{p}_{\mathrm{U}}^{(t)}}\right),\\
		\mathbf{C}_{\mathbf{p}_{\mathrm{U}}^{(t)}\rightarrow\varphi_{i}^{(t)}}^{-1}&= \mathbf{C}_{\mathcal{G}_{\backslash i}^{(t)}}^{-1} + \mathbf{C}_{\psi^{(t)}\rightarrow\mathbf{p}_{\mathrm{U}}^{(t)}}^{-1}.
	\end{align}
\end{subequations}
\subsubsection{Messages from $\varphi_{i}^{(t)}$ to $\theta_{\mathrm{U},i}^{(t)}$}For $\forall t$, $1 \le i\le K$, the message from factor node $\varphi_{i}^{(t)}$ to variable node $\theta_{\mathrm{U},i}^{(t)}$ is given by 
\begin{equation}
	\label{17}
	\Delta _{\varphi_{i}^{(t)}\rightarrow \theta_{\mathrm{U},i}^{(t)}}(\theta_{\mathrm{U},i}^{(t)}) \propto \int_{\mathbf{p}_{\mathrm{U}}^{(t)}}{p(\theta_{\mathrm{U},i}^{(t)}|\mathbf{p}_{\mathrm{U}}^{(t)}) \Delta _{\mathbf{p}_{\mathrm{U}}^{(t)}\rightarrow\varphi_{i}^{(t)} }(\mathbf{p}_{\mathrm{U}}^{(t)})}.
\end{equation}
The integral in \eqref{17} does not have a closed-form expression. To facilitate the subsequent message passing, we approximate this message by a VM distribution based on the far-field assumption as
\begin{figure}
	\centering
	\resizebox{7cm}{!}{\includegraphics{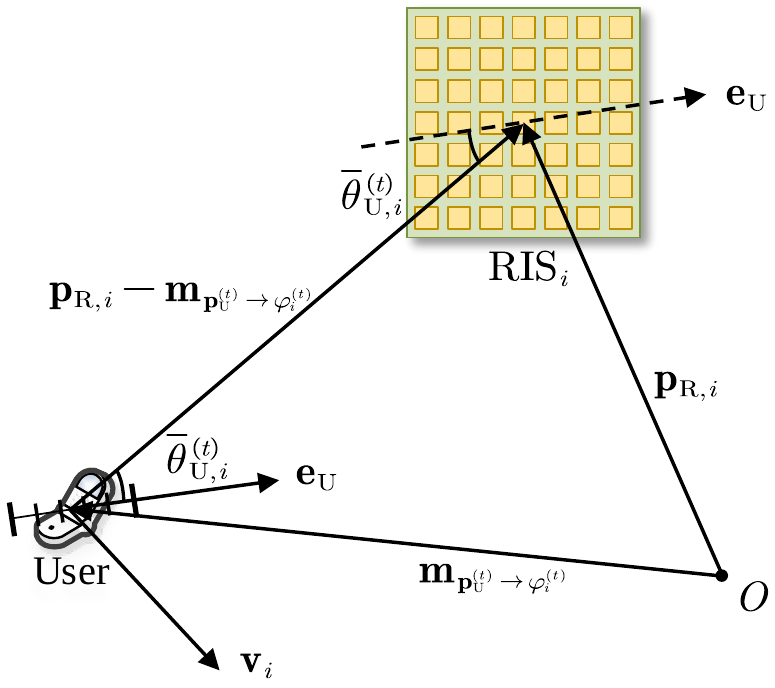}}
	\caption{The geometric relationship for parameters in the VM approximation of message $\Delta _{\varphi_{i}^{(t)}\rightarrow \theta_{\mathrm{U},i}^{(t)}}(\theta_{\mathrm{U},i}^{(t)})$, where $O$ is the origin of coordinates, $\mathbf{P}_{\mathrm{R},i}$ and $\mathbf{m}_{\mathbf{p}_{\mathrm{U}}^{(t)}\rightarrow\varphi_{i}^{(t)}}$ are the vectors point to the $i$-th RIS and the mean user position respectively.}
	\label{fig_VM_approximate}
\end{figure}
\begin{equation}
    \label{VM_appro}
    \Delta _{\varphi_{i}^{(t)}\rightarrow \theta_{\mathrm{U},i}^{(t)}}(\theta_{\mathrm{U},i}^{(t)})\propto\mathcal{M}\left(\pi\theta_{\mathrm{U},i}^{{(t)}};\mu_{\varphi_{i}^{(t)}\rightarrow \theta_{\mathrm{U},i}^{(t)}},\kappa_{\varphi_{i}^{(t)}\rightarrow \theta_{\mathrm{U},i}^{(t)}}\right),
\end{equation}
with
\begin{subequations}
\label{VM_appro_para}
	\begin{align}
		\mu_{\varphi_{i}^{(t)}\rightarrow \theta_{\mathrm{U},i}^{(t)}} &= \pi\bar{\theta} _{\mathrm{U},i}^{(t)},\\
		\kappa_{\varphi_{i}^{(t)}\rightarrow \theta_{\mathrm{U},i}^{(t)}} &=\frac{d_i^2}{\pi^2\left(1-(\bar{\theta} _{\mathrm{U},i}^{(t)})^2\right)\mathbf{v}_i^\mathrm{T}\mathbf{C}_{\mathbf{p}_{\mathrm{U}}^{(t)}\rightarrow\varphi_{i}^{(t)}}\mathbf{v}_i}.
	\end{align}
\end{subequations}
In \eqref{VM_appro_para}, $\mathbf{v}_i$ is the unit directional vector perpendicular to $(\mathbf{p}_{\mathrm{R},i}-\mathbf{m}_{\mathbf{p}_{\mathrm{U}}^{(t)}\rightarrow\varphi_{i}^{(t)}})$ and is in the plane spanned by $\mathbf{e}_{\mathrm{U}}$ and $(\mathbf{p}_{\mathrm{R},i}-\mathbf{m}_{\mathbf{p}_{\mathrm{U}}^{(t)}\rightarrow\varphi_{i}^{(t)}})$; $\bar{\theta} _{\mathrm{U},i}^{(t)}$ is the mean AoA; $d_i$ is the Euclidean distance between $\mathbf{p}_{\mathrm{R},i}$ and $\mathbf{m}_{\mathbf{p}_{\mathrm{U}}^{(t)}\rightarrow\varphi_{i}^{(t)}}$ as shown in Fig. \ref{fig_VM_approximate}. $\mathbf{v}_i$, $\bar{\theta} _{\mathrm{U},i}^{(t)}$ and $d_i$ are given respectively by
\begin{subequations}
	\begin{align}
		\mathbf{v}_i &= \frac{\left( (\mathbf{m}_{\mathbf{p}_{\mathrm{U}}^{(t)}\rightarrow\varphi_{i}^{(t)}}-\mathbf{p}_{\mathrm{R},i})\times \mathbf{e}_{\mathrm{U}} \right)\times\left(\mathbf{m}_{\mathbf{p}_{\mathrm{U}}^{(t)}\rightarrow\varphi_{i}^{(t)}}-\mathbf{p}_{\mathrm{R},i}\right)}{\left\|\left( (\mathbf{m}_{\mathbf{p}_{\mathrm{U}}^{(t)}\rightarrow\varphi_{i}^{(t)}}-\mathbf{p}_{\mathrm{R},i})\times \mathbf{e}_{\mathrm{U}} \right)\times\left(\mathbf{m}_{\mathbf{p}_{\mathrm{U}}^{(t)}\rightarrow\varphi_{i}^{(t)}}-\mathbf{p}_{\mathrm{R},i}\right)\right\|_2},\\
		\bar{\theta}_{\mathrm{U},i}^{(t)}&=\frac{\left( \mathbf{p}_{\mathrm{R},i}-\mathbf{m}_{\mathbf{p}_{\mathrm{U}}^{(t)}\rightarrow \varphi_{i}^{(t)}} \right) ^{\mathrm{T}}\mathbf{e}_{\mathrm{U}}}{\left\| \mathbf{p}_{\mathrm{R},i}-\mathbf{m}_{\mathbf{p}_{\mathrm{U}}^{(t)}\rightarrow \varphi_{i}^{(t)}} \right\|_2},\\
		d_i &= \left\|\mathbf{p}_{\mathrm{R},i}-\mathbf{m}_{\mathbf{p}_{\mathrm{U}}^{(t)}\rightarrow \varphi_{i}^{(t)}}\right\|_2,
	\end{align}
\end{subequations}
where $\times$ denotes the cross product. The derivation of \eqref{VM_appro_para} can be found in Appendix \ref{VMapproximation_derivation}.\par
\subsection{Output Estimations at Time Slot $t$}
The mean of the Gaussian message $\Delta _{\mathbf{p}_{\mathrm{U}}^{(t)}\rightarrow\psi^{(t+1)} }(\mathbf{p}_{\mathrm{U}}^{(t)})$ in \eqref{22}, denoted by $\mathbf{m}_{\mathbf{p}_{\mathrm{U}}^{(t)}\rightarrow\psi^{(t+1)}}$, is used as an output estimate of user position $\mathbf{p}_{\mathrm{U}}^{(t)}$. For the estimation of the user AoAs, we express the message at variable node $\theta_{\mathrm{U},i}^{(t)}$ as
\begin{equation}
	\label{43}
	\Delta _{\theta_{\mathrm{U},i}^{(t)}}(\theta_{\mathrm{U},i}^{(t)})\propto\Delta_{p(\mathbf{y}^{{(t)}}|\boldsymbol{\theta}^{(t)},\boldsymbol{\rho}^{(t)})\rightarrow\theta_{\mathrm{U},i}^{(t)}}(\theta_{\mathrm{U},i}^{(t)}){\Delta _{\varphi_{i}^{(t)}\rightarrow \theta_{\mathrm{U},i}^{(t)}}(\theta_{\mathrm{U},i}^{(t)})}.
\end{equation}
From \eqref{22_n}-\eqref{36}, $\Delta _{\theta_{\mathrm{U},i}^{(t)}}(\theta_{\mathrm{U},i}^{(t)})$ is approximated as a VM distribution. We obtain an output estimate of $\theta_{\mathrm{U},i}^{(t)}$ as $\mu_{\theta_{\mathrm{U},i}^{(t)}}/\pi$, where $\mu_{\theta_{\mathrm{U},i}^{(t)}}$ is the mean direction of $\Delta _{\theta_{\mathrm{U},i}^{(t)}}(\theta_{\mathrm{U},i}^{(t)})$.
%
\subsection{Overall Algorithm}
\begin{algorithm}[t]
	\caption{BULT Algorithm} 
	\label{tracking_algorithm} 
	{\bf Input:} Observed signal $\mathbf{y}^{(1)}, \dots, \mathbf{y}^{(t_0)}$, initial user position estimation $\mathbf{m}_{\mathbf{p}_{\mathrm{U}}^{(0)}\rightarrow\psi^{(1)}}$ and with its covariance matrix $\mathbf{C}_{\mathbf{p}_{\mathrm{U}}^{(0)}\rightarrow\psi^{(1)}}$.
	\\
	{\bf Output:} User position estimations $\{\mathbf{m}_{\mathbf{p}_{\mathrm{U}}^{(t)}\rightarrow\psi^{(t+1)}}\}$, user AoA estimations $\{\mu_{\theta_{\mathrm{U},i}^{(t)}}/\pi\}$, and equivalent path gain estimations $\{\hat{\rho}^{(t)}_i\}$ for $1\leq i \leq K$, $1\leq t \leq t_0$.
	\begin{algorithmic}[1] 
		\FOR{$t = 1$ to $t_0$}
		\STATE{Calculate  $\mathbf{m}_{\psi^{(t)}\rightarrow\mathbf{p}_{\mathrm{U}}^{(t)}}$
			and $\mathbf{C}_{\psi^{(t)}\rightarrow\mathbf{p}_{\mathrm{U}}^{(t)}}$ by \eqref{33_nn}.}
		\STATE {\textbf{Initialization:} for $\forall i$, $\mathbf{m}_{\mathcal{G}_{\backslash i}^{(t)}} =\mathbf{m}_{\psi^{(t)}\rightarrow\mathbf{p}_{\mathrm{U}}^{(t)}}$, $\mathbf{C}_{\mathcal{G}_{\backslash i}^{(t)}} =\mathbf{C}_{\psi^{(t)}\rightarrow\mathbf{p}_{\mathrm{U}}^{(t)}}$.}
		\REPEAT
		\STATE {For $\forall i$, update the message $\Delta _{\mathbf{p}_{\mathrm{U}}^{(t)}\rightarrow\varphi_{i}^{(t)} }(\mathbf{p}_{\mathrm{U}}^{(t)})$ by \eqref{29}.}
		\STATE {For $\forall i$, update the message $\Delta _{\varphi_{i}^{(t)}\rightarrow \theta_{\mathrm{U},i}^{(t)}}(\theta_{\mathrm{U},i}^{(t)})$ by \eqref{VM_appro}, \eqref{VM_appro_para}.}
		\STATE {For $\forall i$, update the posterior pdf estimation $p(\theta_{\mathrm{U},i}^{(t)}|\mathbf{y}^{(t)})$ and equivalent path gain estimation $\hat{\rho}^{(t)}_i$ by the VALSE algorithm}.
		\STATE {For $\forall i$, update the message $\Delta _{\theta_{\mathrm{U},i}^{(t)}\rightarrow \varphi_{i}^{(t)}}(\theta_{\mathrm{U},i}^{(t)})$ by \eqref{36}, \eqref{37}.}
		\STATE {For $\forall i$, update $\mathbf{m}_{\mathcal{G}_{\backslash i}^{(t)}}$ and $\mathbf{C}_{\mathcal{G}_{\backslash i}^{(t)}}$ by \eqref{36_nn}.}
		\UNTIL stopping criterion
		\STATE {Calculate $\mathbf{m}_{\mathbf{p}_{\mathrm{U}}^{(t)}\rightarrow\psi^{(t+1)}}$ and $\mathbf{C}_{\mathbf{p}_{\mathrm{U}}^{(t)}\rightarrow\psi^{(t+1)}}$ by and \eqref{32_nn}.}
		\ENDFOR
		\STATE \textbf{return} $\{\mathbf{m}_{\mathbf{p}_{\mathrm{U}}^{(t)}\rightarrow\psi^{(t+1)}}\}$, $\{\mu_{\theta_{\mathrm{U},i}^{(t)}}/\pi\}$, and $\{\hat{\rho}^{(t)}_i\}$ for $1\leq t \leq t_0$, $1\leq i \leq K$.
	\end{algorithmic}
\end{algorithm}
Based on the discussions in the preceding subsections, our BULT algorithm is summarized in Algorithm \ref{tracking_algorithm}. Algorithm \ref{tracking_algorithm} is an online algorithm that estimates the user position $\mathbf{p}_{\mathrm{U}}^{(t)}$ and AoAs $\{\theta_{\mathrm{U},i}^{(t)}\}$ jointly in each time slot by iteratively passing messages between the position tracking module and the AoA estimation module. Specifically, lines 5-6 of Algorithm \ref{tracking_algorithm} correspond to the calculation of message passing from the position tracking module to the AoA estimation module. Lines 7-8 correspond to the calculation of message passing from the the AoA estimation module to the position tracking module. Line 2 and line 11 correspond to the calculation of message passing along the Markov chain.
Note that the factor graph in Fig. \ref{fig_factorgraph} is loopy, so the convergence of the BULT algorithm cannot be guaranteed. As inspired by \cite{parker2014bilinear}, we apply the damping technique on the messages $\{\Delta _{\mathbf{p}_{\mathrm{U}}^{(t)}\rightarrow\varphi_{i}^{(t)} }(\mathbf{p}_{\mathrm{U}}^{(t)})\}$ to improve the convergence.\par
As for the initialization in the first time slot $t=1$, we assume that a coarse user position information in the form of mean $\mathbf{m}_{\mathbf{p}_{\mathrm{U}}^{(0)}\rightarrow\psi^{(1)}}$ and covariance matrix $\mathbf{C}_{\mathbf{p}_{\mathrm{U}}^{(0)}\rightarrow\psi^{(1)}}$ is available, such information can be provided, e.g., by the Global Positioning System (GPS). Alternatively, we can run the BULT algorithm without any prior position information (e.g., by setting $\mathbf{m}_{\mathbf{p}_{\mathrm{U}}^{(0)}\rightarrow\psi^{(1)}}=\mathbf{0}$ and $\mathbf{C}_{\mathbf{p}_{\mathrm{U}}^{(0)}\rightarrow\psi^{(1)}}=\eta\mathbf{I}$ where $\eta$ is a large positive number). In this case, the estimation of $\{\theta_{\mathrm{U},i}^{(1)}\}_{i=1}^{K}$ given by the AoA estimation module are unordered which causes the mismatch problem between the variable nodes $\{\theta_{\mathrm{U},i}^{(1)}\}_{i=1}^{K}$ and the factor nodes $\{\varphi_{i}^{(1)}\}_{i=1}^{K}$. To obtain a correct match, we exhaustively search all the possible matching schemes and select the one with the minimum $\mathrm{Tr}(\mathbf{C}_{\mathbf{p}_{\mathrm{U}}^{(1)}\rightarrow\psi^{(2)}})$.
\subsection{Complexity of BULT}
	The complexity of the BULT algorithm mainly arises from the variational Bayesian method and the Gaussian message approximation. For any time slot, the complexity of the variational Bayesian method according to \cite{badiu2017variational} is $\mathcal{O}(n_1K^3N_{\mathrm{U}}+n_1KN_{\mathrm{U}}^2)$, where $n_1$ is the number of iterations in the variational Bayesian method. The complexity of Gaussian message approximation is $\mathcal{O}(n_2K)$, where $n_2$ is the iteration number of GDM in \eqref{64}. Therefore, the overall complexity of BULT is given by $\mathcal{O}(n_3n_1K^3N_{\mathrm{U}}+n_3n_1KN_{\mathrm{U}}^2+n_3n_2K)$, where $n_3$ is the iteration number of message passing between the position tracking module and the AoA estimation module, which depends on the stopping criterion and is empirically less than 15.
\section{Bayesian Cram\'er Rao Bound}
\label{Section4}
In this section, we derive the BCRB for the estimation of the user position and AoAs over time. The BCRB acts as a benchmark of our tracking algorithm. Considering the geometric constraint in \eqref{eq10}, we take a parameter set in the time slot $t$ as $\boldsymbol{\gamma }^{(t)}=[ (\mathbf{p}_\mathrm{U}^{(t)})^\mathrm{T},(\angle{\boldsymbol{\rho}}^{(t)})^\mathrm{T},(\vert\boldsymbol{\rho}^{(t)}\vert)^\mathrm{T}]^\mathrm{T} \in \mathbb{R}^{2K+3}$, where the $i$-th term of $\angle{\boldsymbol{\rho}}^{(t)}$ and $\vert\boldsymbol{\rho}^{(t)}\vert$ are $\angle{{\rho}_{i}^{(t)}}$ and $\vert {\rho}_{i}^{(t)} \vert $, respectively represent the angle and amplitude of the equivalent complex path gain ${\rho}_{i}^{(t)}$. In the sequel, we first derive the Fisher information matrix (FIM) in a single time slot by neglecting the probability model and regarding $\boldsymbol{\gamma }^{(t)}$ as a deterministic parameter set. We then compute the BCRB based on the derived FIM. \par
\subsection{FIM Calculation for a Single Time Slot}
%
Consider the problem of estimating a deterministic parameter set $\boldsymbol{\gamma }^{(t)}$ in time slot $t$ without any prior information from time slot $t-1$. The MSE of the unbiased estimator is lower bounded by the Cram\'er Rao bound (CRB) which is the diagonal of the inverse of FIM $\mathbf{J}^{(t)}\in\mathbb{R}^{(2K+3)\times(2K+3)}$. 
For the signal model in \eqref{eq9}, the $(i,j)$-th entry of $\mathbf{J}^{(t)}$ can be calculated according to the following lemma \cite{collier2005fisher}:
\begin{lemma}
	For an N-dimensional complex Gaussian noise disturbed observation signal $\mathbf{y}^{(t)} \sim \mathcal{CN}(\mathbf{y}^{(t)};\boldsymbol{\mu}^{(t)},\boldsymbol{\Sigma}^{(t)})$, the $(i,j)$-th entry of the FIM is given by
	\begin{align}
		[\mathbf{J}^{(t)}]_{i,j}=&2 \mathcal{R}\left\{\frac{\partial (\boldsymbol{\mu}^{(t)})^{\mathrm{H}}}{\partial \gamma_{i}^{(t)}} \left(\boldsymbol{\Sigma}^{(t)}\right)^{-1} \frac{\partial \boldsymbol{\mu}^{(t)}}{\partial \gamma_{j}^{(t)}}\right\}\notag\\
		&+\mathrm{Tr} \left\{\left(\boldsymbol{\Sigma}^{(t)}\right)^{-1} \frac{\partial \boldsymbol{\Sigma}^{(t)}}{\partial \gamma_{i}^{(t)}} \left(\boldsymbol{\Sigma}^{(t)}\right)^{-1} \frac{\partial \boldsymbol{\Sigma}^{(t)}}{\partial \gamma_{j}^{(t)}}\right\}.
	\end{align}
\end{lemma}
In our scenario, we have $\boldsymbol{\mu}^{(t)} =\sum_{i=1}^K{\rho _{i}^{(t)}\mathbf{a}_{\mathrm{U}}(\theta_{\mathrm{U},i}^{{(t)}})}$ and $\boldsymbol{\Sigma}^{(t)}=\sigma_n^2\mathbf{I}$. Since $\boldsymbol{\Sigma}^{(t)}$ is independent with the parameter set $\boldsymbol{\gamma}^{(t)}$, the FIM of $\boldsymbol{\gamma}^{(t)}$ is given as
\begin{align}
	\label{FIMcal}
	\mathbf{J}^{(t)}  =\frac{2}{\sigma_n ^2}\!\sum_{n=1}^{N_\mathrm{U}}\!{\left(\!\frac{\partial \mathcal{R}\!\{{\mu}_{n}^{(t)} \!\}}{\partial \boldsymbol{\gamma }^{(t)}}\!\left(\!\frac{\partial \mathcal{R}\!\{ {\mu}_{n}^{(t)} \!\}}{\partial \boldsymbol{\gamma }^{(t)}}\!\right) ^\mathrm{T}\!+\!\frac{\partial \mathcal{I}\!\{ {\mu}_{n}^{(t)} \!\}}{\partial \boldsymbol{\gamma }^{(t)}}\left(\!\frac{\partial \mathcal{I}\!\{ {\mu}_{n}^{(t)} \!\}}{\partial \boldsymbol{\gamma }^{(t)}} \!\right)^\mathrm{T}\right)},
\end{align}
where ${\mu}_{n}^{(t)}$ is the $n$-th entry of $\boldsymbol{\mu}^{(t)}$ given by
\begin{equation}
	{\mu}_{n}^{(t)}=\sum_{i=1}^K{\vert {\rho}_{i}^{(t)} \vert e^{j(\pi(n-1)\theta_{\mathrm{U},i}^{{(t)}}+\angle{{\rho}_{i}^{(t)}})}}.
\end{equation}
For $1\leq n\leq N_{\mathrm{U}}$ and $1\leq i \leq K$, we have 
\begin{subequations}
\label{44}
	\begin{align}
		\frac{\partial \mathcal{R}\!\left\{{\mu}_{n}^{(t)} \!\right\}}{\partial \mathbf{p}_\mathrm{U}^{(t)}} &\!=\!\sum_{i=1}^K\!{\left(\!-\vert{\rho}_{i}^{(t)} \vert\pi\!\left(\!n \!-\! 1\!\right)\!\sin\!\left(\! \pi\!\left(\!n \!-\! 1\!\right)  \theta_{\mathrm{U},i}^{{(t)}}\! +\! \angle{{\rho}_{i}^{(t)}} \right)\!\right)\! \frac{\partial \theta_{\mathrm{U},i}^{{(t)}}}{\partial \mathbf{p}_{\mathrm{U}}^{(t)}}},\label{partial1}\\
		\frac{\partial \mathcal{I}\!\left\{{\mu}_{n}^{(t)} \!\right\}}{\partial \mathbf{p}_\mathrm{U}^{(t)}} &\!=\! \sum_{i=1}^K\!{\left(\!\vert {\rho}_{i}^{(t)} \vert\pi\!\left(\!n \!-\! 1\! \right)\!  \cos\!  \left(\! \pi\!  \left(\! n\! -\! 1\! \right)  \theta_{\mathrm{U},i}^{{(t)}}\! +\! \angle{{\rho}_{i}^{(t)}} \right)\! \right)\! \frac{\partial \theta_{\mathrm{U},i}^{{(t)}}}{\partial \mathbf{p}_{\mathrm{U}}^{(t)}}},\label{partial2}
	\end{align}
\end{subequations}
where $\frac{\partial \theta_{\mathrm{U},i}^{{(t)}}}{\partial \mathbf{p}_\mathrm{U}^{(t)}}$ is calculated according to \eqref{eq10} given by
\begin{equation}
	\frac{\partial \theta_{\mathrm{U},i}^{{(t)}}}{\partial \mathbf{p}_\mathrm{U}^{(t)}}=  \frac{-\mathbf{e}_\mathrm{U}+({\mathbf{e}_{i}^{(t)}})^{\mathrm{T}}\mathbf{e}_\mathrm{U}\mathbf{e}_{i}^{(t)}}{\left\| \mathbf{p}_{\mathrm{R},i}-\mathbf{p}_\mathrm{U}^{(t)} \right\| _2},\label{geopartial}
\end{equation} 
with  $\mathbf{e}_{i}^{(t)}=\frac{\left( \mathbf{p}_{\mathrm{R},i}-\mathbf{p}_\mathrm{U}^{(t)} \right)}{\left\| \mathbf{p}_{\mathrm{R},i}-\mathbf{p}_{\mathrm{U}}^{(t)} \right\| _2}$. Similarly, we have
\begin{subequations}
\label{46_n}
	\begin{align}
		\frac{\partial \mathcal{R} \left\{ {\mu}_{n}^{(t)} \right\}}{\partial \angle{{\rho}_{i}^{(t)}}}&=-\vert {\rho}_{i}^{(t)} \vert\sin \left(\pi \left( n-1 \right) \theta_{\mathrm{U},i}^{{(t)}}+\angle{{\rho}_{i}^{(t)}} \right), \label{partial5}\\
		\frac{\partial \mathcal{I} \left\{ {\mu}_{n}^{(t)} \right\}}{\partial \angle{{\rho}_{i}^{(t)}}}&=\vert {\rho}_{i}^{(t)} \vert\cos \left(\pi \left( n-1 \right) \theta_{\mathrm{U},i}^{{(t)}}+\angle{{\rho}_{i}^{(t)}} \right),\label{partial6} \\
		\frac{\partial \mathcal{R} \left\{{\mu}_{n}^{(t)} \right\}}{\partial \vert {\rho}_{i}^{(t)} \vert}&=\cos  \left(\pi \left( n-1 \right) \theta_{\mathrm{U},i}^{{(t)}}+\angle{{\rho}_{i}^{(t)}} \right), \label{partial3}\\
		\frac{\partial \mathcal{I} \left\{ {\mu}_{n}^{(t)} \right\}}{\partial \vert {\rho}_{i}^{(t)} \vert}&=\sin  \left(\pi \left( n-1 \right) \theta_{\mathrm{U},i}^{{(t)}}+\angle{{\rho}_{i}^{(t)}} \right). \label{partial4}
	\end{align}
\end{subequations}
By substituting \eqref{44}-\eqref{46_n} into \eqref{FIMcal}, we obtain the FIM in the current time slot without considering the prior information from the preceding time slot.\par
\subsection{BCRB for Successive Time Slots}
The BCRB is derived based on the Bayesian Fisher information matrix (BFIM). For an arbitrary time slot $t$, the BFIM of $\boldsymbol{\gamma}^{(t)}$ denoted by $\mathbf{J}_\mathrm{B}^{(t)}$ is recursively calculated by \cite{tichavsky1998posterior}
\begin{align}
	\label{46}
	\mathbf{J}_\mathrm{B}^{(t)} &= \mathbf{J}^{(t)} +\mathbf{G}_{22}^{(t)}-\mathbf{G}_{21}^{(t)}( \mathbf{J}_\mathrm{B}^{(t-1)} + \mathbf{G}_{11}^{(t)})^{-1}\mathbf{G}_{12}^{(t)},
\end{align}
where $\mathbf{J}^{(t)}$ is given by \eqref{FIMcal} and the else items in \eqref{46} are defined as
\begin{subequations}
	\begin{align}
		\mathbf{G}_{11}^{(t)} &= \mathbb{E}\left[-\frac{\partial^2\log p(\boldsymbol{\gamma}^{(t)}|\boldsymbol{\gamma}^{(t-1)})}{\partial\boldsymbol{\gamma}^{(t-1)}\partial(\boldsymbol{\gamma}^{(t-1)})^{\mathrm{T}}}\right], \\
		\mathbf{G}_{12}^{(t)} &= \mathbb{E}\left[-\frac{\partial^2\log p(\boldsymbol{\gamma}^{(t)}|\boldsymbol{\gamma}^{(t-1)})}{\partial\boldsymbol{\gamma}^{(t-1)}\partial(\boldsymbol{\gamma}^{(t)})^{\mathrm{T}}}\right], \\
		\mathbf{G}_{21}^{(t)} &= \mathbb{E}\left[-\frac{\partial^2\log p(\boldsymbol{\gamma}^{(t)}|\boldsymbol{\gamma}^{(t-1)})}{\partial\boldsymbol{\gamma}^{(t)}\partial(\boldsymbol{\gamma}^{(t-1)})^{\mathrm{T}}}\right], \\
		\mathbf{G}_{22}^{(t)} &= \mathbb{E}\left[-\frac{\partial^2\log p(\boldsymbol{\gamma}^{(t)}|\boldsymbol{\gamma}^{(t-1)})}{\partial\boldsymbol{\gamma}^{(t)}\partial(\boldsymbol{\gamma}^{(t)})^{\mathrm{T}}}\right].
		\end{align}
	\end{subequations}
Due to the independence of $\boldsymbol{\rho}^{(t)}$ in different time slots, we obtain $p(\boldsymbol{\gamma }^{(t)}|\boldsymbol{\gamma }^{(t-1)})=p(\mathbf{p}_{\mathrm{U}}^{(t)}|\mathbf{p}_{\mathrm{U}}^{(t-1)})$. Further calculation yields $	\mathbf{G}_{22}^{(t)} = \mathbf{G}_{11}^{(t)} = -\mathbf{G}_{12}^{(t)} = -\mathbf{G}_{21}^{(t)} \in \mathbb{R}^{(2K+3)\times(2K+3)}$, where
\begin{equation}
		\mathbf{G}_{11}^{(t)}=\left[ \begin{matrix}
		\mathbb{E}\left[-\frac{\partial^2\log p\left( \mathbf{p}_\mathrm{U}^{(t+1)}|\mathbf{p}_\mathrm{U}^{(t)} \right)}{\partial\mathbf{p}_\mathrm{U}^{(t)}\partial(\mathbf{p}_\mathrm{U}^{(t)})^{\mathrm{T}}} \right]&		\mathbf{0}\\
		\mathbf{0}&		\mathbf{0}\\
	\end{matrix} \right],
\end{equation}
with $\mathbb{E}\left[-\frac{\partial^2\log p\left( \mathbf{p}_\mathrm{U}^{(t+1)}|\mathbf{p}_\mathrm{U}^{(t)}\right)}{\partial\mathbf{p}_\mathrm{U}^{(t)}\partial(\mathbf{p}_\mathrm{U}^{(t)})^{\mathrm{T}}} \right]=\mathbf{C}_\mathrm{q}^{-1}$. By initializing ${\mathbf{J}}_\mathrm{B}^{(0)}=\mathbb{E}[-\frac{\partial^2\log p(\mathbf{p}_\mathrm{U}^{(0)})}{\partial\mathbf{p}_\mathrm{U}^{(0)}\partial(\mathbf{p}_\mathrm{U}^{(0)})^{\mathrm{T}}}]$, the BCRB of the estimation of $ \hat{\boldsymbol{\gamma}}^{(t)}$ in the $t$-th time slot is given by
\begin{equation}
	\mathbb{E}\left[\![\!(\!\hat{\boldsymbol{\gamma}}^{(\!t\!)}\!-\!\boldsymbol{\gamma}^{(\!t\!)}\!)(\!\hat{\boldsymbol{\gamma}}^{(\!t\!)}\!-\!\boldsymbol{\gamma}^{(\!t\!)}\!)^\mathrm{H}]_{i,i}\!\right]\!\geq\![\!(\mathbf{J}_\mathrm{B}^{(\!t\!)}\!)\!^{-\!1}\!]_{i,i},\quad \mathrm{for}\;\forall 1\!\leq \!i\!\leq\!2K\!+3.
\end{equation}
\par
As for the BCRB of AoAs estimation $\boldsymbol{\theta}^{(t)}$, we exploit the theory of parameter transformation for CRLB \cite{kay1993fundamentals}. Assuming a function mapping $\boldsymbol{\theta}^{(t)}=\mathrm{g}_t(\boldsymbol{\gamma}^{(t)})$ associates $\boldsymbol{\gamma}^{(t)} $ and $\boldsymbol{\theta}^{(t)}$, the FIM of $\boldsymbol{\theta}^{(t)}$ denoted by $ \mathbf{J}_{\boldsymbol{\theta}}^{(t)}\in\mathbb{R}^{K\times K}$ is obtained by 
\begin{equation}
	\mathbf{J}_{\boldsymbol{\theta}}^{(t)} = \mathbf{T}_{\boldsymbol{\theta}}^{(t)} \mathbf{J}_\mathrm{B}^{(t)} (\mathbf{T}_{\boldsymbol{\theta}}^{(t)})^\mathrm{T},
\end{equation}
where $\mathbf{T}_{\boldsymbol{\theta}}^{(t)}\in \mathbb{R}^{K\times(2K+3)}$ is the Jacobian matrix of $\mathrm{g}_t(\boldsymbol{\gamma}^{(t)})$ whose entries are obtained by 
\begin{equation}
	\mathbf{T}_{\boldsymbol{\theta }}^{(t)}=\left[ \frac{\partial \boldsymbol{\theta }^{(t)}}{\partial ({\mathbf{p}_{\mathrm{U}}^{(t)}})^{\mathrm{T}}}, \frac{\partial \boldsymbol{\theta}^{(t)}}{\partial (\angle{\boldsymbol{\rho}}^{(t)})^{\mathrm{T}}}, \frac{\partial \boldsymbol{\theta}^{(t)}}{\partial (\vert\boldsymbol{\rho}^{(t)}\vert)^{\mathrm{T}}} \right],
\end{equation}
where $ \frac{\partial \boldsymbol{\theta }_T}{\partial (\mathbf{p}_{\mathrm{U}}^{(t)})^{\mathrm{T}}}$ is given according to \eqref{geopartial} and $\frac{\partial \boldsymbol{\theta}^{(t)}}{\partial (\angle{\boldsymbol{\rho}}^{(t)})^{\mathrm{T}}}=\frac{\partial \boldsymbol{\theta}^{(t)}}{\partial (\vert\boldsymbol{\rho}^{(t)}\vert)^{\mathrm{T}}}= \mathbf{0}$. Similarly, for any $ 1\leq i \leq K$, the BCRB for estimation of $\hat{\boldsymbol{\theta }}^{(t)}$ in the $t$-th time slot is given by
\begin{equation}
	\mathbb{E}\left[\![(\hat{\boldsymbol{\theta }}^{(t)}\!\!-\!\boldsymbol{\theta}^{(t)}\!)(\hat{\boldsymbol{\theta }}^{(t)}\!\!-\!\boldsymbol{\theta}^{(t)}\!)^\mathrm{H}]_{i,i}\!\right]\!\geq\! [(\mathbf{J}_{\boldsymbol{\theta}}^{(t)}\!)^{-1}]_{i,i},\quad \mathrm{for}\; \forall 1\!\leq \!i\!\leq\! K. 
\end{equation}
The above BCRB is useful in the beamforming design discussed in the next section.
\section{Beamforming Design for BS and RISs}
\label{Section5}
The beamforming (BF) of the BS and the passive beamforming (PBF) of the RISs need to be appropriately designed to assist the user tracking. In this section, we first introduce the beamforming design based on the minimization of the BCRB. We then provide a position-based directional beamforming design as a low-complexity solution.
\subsection{BCRB Based Beamforming Design}\label{VA}
In Section \ref{Section4}, we have derived the BCRB of the parameter estimation problem, which provides a metric for evaluating the tracking performance. By optimizing the BF and PBF jointly, we minimize the BCRB of the user position in the considered tracking problem corresponding to the first three diagonal elements of $(\mathbf{J}_{\mathrm{B}}^{(t)})^{-1}$. Since $\boldsymbol{\gamma }^{(t)}$ is unknown, the user position $\mathbf{p}_{\mathrm{U}}^{(t)}$ and equivalent complex path gain $\boldsymbol{\rho}^{(t)}$ are approximated respectively by the estimated user position $\mathbf{m}_{\mathbf{p}_{\mathrm{U}}^{(t-1)}\rightarrow\psi^{(t)}}$ and the vector valued function ${\boldsymbol{\varrho}}^{(t-1)}(\boldsymbol{\omega}^{(t)},\mathbf{f}^{(t)})$, where $\boldsymbol{\omega}^{(t)} \in \mathbb{R}^{KN_\mathrm{R}}$ collects the PBF vector of RISs as $\boldsymbol{\omega}^{(t)}=[\boldsymbol{\omega}_{1}^{(t)},\dots,\boldsymbol{\omega}_{K}^{(t)}]^{\mathrm{T}}$. Referring to \eqref{11}, the $i$-th term of function ${\boldsymbol{\varrho}}^{(t-1)}(\boldsymbol{\omega}^{(t)},\mathbf{f}^{(t)})$ is expressed as
\begin{equation}
	{\varrho}_i^{{(t\!-\!1)}}(\!\boldsymbol{\omega}_i^{(t)},\mathbf{f}^{(t)}\!)\!=\!  { \hat{\rho} _{\mathrm{UB},i}^{{(t\!-\!1)}}\mathbf{a}_{\mathrm{R}}^{\mathrm{H}}(\!\hat{\vartheta} _{\mathrm{R},i}^{{(t\!-\!1)}} \!)}\mathrm{diag}\!\left(\!\boldsymbol{\omega}_i^{(t)}\!\right){\mathbf{a}_{\mathrm{R}}(\!\theta _{\mathrm{R},i}\!) \mathbf{a}_{\mathrm{B}}^{\mathrm{H}}(\!\vartheta _{\mathrm{B},i}\!)\mathbf{f}^{(t)}},
\end{equation}
where $\hat{\vartheta} _{\mathrm{R},i}^{{(t-1)}}$ is the average AoD for the $i$-th RIS at the $(t-1)$-th time slot, i.e.,
\begin{equation}
	\label{54}
	\hat{\vartheta}_{\mathrm{R},i}^{{(t-1)}}=\frac{\left(\mathbf{m}_{\mathbf{p}_{\mathrm{U}}^{(t-1)}\rightarrow\psi^{(t)}}-\mathbf{p}_{\mathrm{R},i}\right)^{\mathrm{T}}\mathbf{e}_{\mathrm{R},i}}{\Vert \mathbf{m}_{\mathbf{p}_{\mathrm{U}}^{(t-1)}\rightarrow\psi^{(t)}}-\mathbf{p}_{\mathrm{R},i} \Vert_2},
\end{equation}
with $\mathbf{e}_{\mathrm{R},i}$ being the known direction vector of the $i$-th RIS, and
\begin{equation}
	\hat{\rho}_{\mathrm{UB},i}^{{(t-1)}}=\frac{\hat{\rho}_i^{(t-1)}}{{ \mathbf{a}_{\mathrm{R}}^{\mathrm{H}}(\hat{\vartheta} _{\mathrm{R},i}^{{(t-1)}} )}\mathrm{diag}\left(\hat{\boldsymbol{\omega}}_i^{{(t-1)}}\right){\mathbf{a}_{\mathrm{R}}(\theta _{\mathrm{R},i}) \mathbf{a}_{\mathrm{B}}^{\mathrm{H}}( \vartheta _{\mathrm{B},i} )\hat{\mathbf{f}}^{(t-1)}}},
\end{equation}
where $\hat{\rho}_i^{{(t-1)}}$ is the estimated equivalent complex path gain, $\hat{\boldsymbol{\omega}}_i^{(t-1)}$ and $\hat{\mathbf{f}}^{(t-1)}$ are the beamforming vectors in time slot $t-1$. Therefore, the optimization problem in time slot $t$ is formulated as
\begin{mini*}|s|
	{\substack{\boldsymbol{\omega }^{(t)},\mathbf{f}^{(t)}}}
	{\sum_{j=1}^3\!{\left[\!\left( \mathbf{J}_{\mathrm{B}}^{(t)}\!\left(\! \mathbf{m}_{\mathbf{p}_{\mathrm{U}}^{(t\!-\!1)}\!\rightarrow\psi^{(t)}},\boldsymbol{{\varrho}}^{(t\!-\!1)}(\boldsymbol{\omega }^{(t)}\!,\mathbf{f}^{(t)})\!\right)\! \right) ^{-1}\!\right]_{j,j}}}
	{}
	{\text{(P1)}:}
	\addConstraint{| \boldsymbol{\omega }_{i,j}^{(t)} |=1,\mathrm{for}\quad 1\leq i\leq K ,\; 1\leq j\leq N_\mathrm{U}}
	\addConstraint{\| \mathbf{f}^{\left( t \right)} \| =1.}
\end{mini*}
The optimal solution of (P1) is difficult to be found due to the non-convexity of the objective function and the constraints. We propose to use the \textit{alternating gradient descent method} (AGDM) which alternately optimizes the PBF of the RISs and the BF of the BS to obtain a locally optimal solution.
\subsection{Directional Beamforming Design}
The complexity of the beamforming design in \ref{VA} is high due to the high-dimension of the optimization variables and the iterative nature of the design algorithm. We next give the position based directional BF and PBF design as a low-complexity solution.\par
For the passive beamforming design, the PBF for the $i$-th RIS in time slot $t$ is given by
\begin{equation}
	\label{dir_PBF}
	\boldsymbol{\omega}_i^{{(t)}}=\mathbf{a}_{\mathrm{R}}( \hat{\vartheta}_{\mathrm{R},i}^{{(t-1)}})\odot\mathbf{a}_{\mathrm{R}}^{\mathrm{H}}( \theta _{\mathrm{R},i}),
\end{equation}
which aligns the reflect beam of the $i$-th RIS to the estimation of user position $\mathbf{m}_{\mathbf{p}_{\mathrm{U}}^{(t-1)}\rightarrow\psi^{(t)}}$. Thus we obtain the closed-form solution of PBF. Similar RIS PBF designs can be found in \cite{hu2020location,cai2021hierarchical}.\par
For the beamforming design of the BS, as the positions of the RISs are fixed and known, a concise and effective BS beamforming design is given as 
\begin{equation}
	\label{BS_BF}
	\mathbf{f}^{(t)}=\sum_{i=1}^{K}{w_i^{(t)}\mathbf{a}_{\mathrm{B}}(\vartheta _{\mathrm{B},i})},
\end{equation}
where $w_i^{(t)}$ is the weight of the $i$-th directional beam obtained by solving the following optimization problem:
\begin{mini*}|s|
	{\substack{\boldsymbol{w}^{(t)}}}
	{\sum_{j=1}^3{\left[ \left( \mathbf{J}_{\mathrm{B}}^{(t+1)}\left( \mathbf{\hat{p}}_{\mathrm{U}}^{(t)},\boldsymbol{{\varrho}}^{(t)}(\boldsymbol{w}^{\left( t \right)}) \right) \right) ^{-1} \right] _{j,j}}}
	{}
	{\text{(P2)}:}
	\addConstraint{\left\|{\sum_{i=1}^{K}{w_i^{(t)}\mathbf{a}_{\mathrm{B}}(\vartheta _{\mathrm{B},i})}}\right\|_2 =1,}
\end{mini*}
where $\boldsymbol{w}^{(t)} = [w_1^{(t)},\dots,w_K^{(t)}]^{\mathrm{T}}$. Here we also use the GDM to obtain a locally optimal solution of problem (P2). In the angle domain, the beamforming design in \eqref{BS_BF} provides multiple beam lobes with each aligned to a different RIS.
\section{Numerical Experiments}
\label{Section6}
In this section, we conduct numerical experiments to demonstrate the performance of the BULT algorithm and the beamforming design in different scenarios.
\begin{table}[t]
	\centering
	\small
	\caption{System Parameters}
	\begin{tabular}{p{\columnwidth/5}p{\columnwidth*4/6}}
		\hline
		\textbf{Parameter}&\textbf{Value}\\
		\hline
		 $f_c$&$28$ GHz\\
		$\sigma_n^2$&$-84$ dBm\\
		$N_\mathrm{B}$&32\\
		$N_\mathrm{R}$&32, 64, 96\\
		$N_\mathrm{U}$&17\\
		$\mathbf{p}_\mathrm{B}$&(20 m, 0 m, 0 m)\\
		$\mathbf{p}_{\mathrm{R},i}$&(-35 m, 5 m, -10 m), (-30 m, 20 m, 10 m), (-20 m, 25 m, 20 m), (-10 m, 40 m, 10 m), (0 m, 20 m, 10 m), (10 m, 15 m, 20 m), (30 m, 20 m, 5 m)\\
		$\mathbf{e}_\mathrm{R,i}$&(0, 1, 0), (1, 0, 0), (1, 0, 0), (1, 0, 0), (1, 0, 0), (1, 0, 0), (0, 1, 0)\\
		$\mathbf{e}_\mathrm{U}$&(1, 0, 0)\\
		\hline
	\end{tabular}
	\label{table1}
\end{table}
\subsection{System Parameters and Performance Metric}
\label{VI_subsectionA}
We deploy the RISs and the BS appropriately to ensure the far-field assumption used in \eqref{eq8} holds. The parameter $\rho_{\mathrm{UB},i}^{(t)}$ in the equivalent complex path gains is generated by following \cite{wang2021joint,basar2019wireless} as
\begin{equation}
	\label{equivalent_pathgain}
	\rho_{\mathrm{UB},i}^{(t)} = \frac{\lambda}{4 \pi\left(d_{\mathrm{B},\mathrm{R}_i}+d_{\mathrm{R}_i,\mathrm{U}}\right)} e^{-j \frac{2\pi}{\lambda}(d_{\mathrm{B},\mathrm{R}_i}+d_{\mathrm{R}_i,\mathrm{U}})},
\end{equation}
where $d_{\mathrm{B},\mathrm{R}_i}$ and $d_{\mathrm{R}_i,\mathrm{U}}$ are the distances from the BS to the $i$-th RIS and the $i$-th RIS to the user respectively. We use the directional beamforming design for $\boldsymbol{\omega}^{(t)}$ and $\mathbf{f}^{(t)}$, unless otherwise specified. The system parameter settings are listed in Table \ref{table1}.\par
We assume that the tracking interval is $20$ ms, which makes it possible to track the user in a high-speed mobile scenario. The user is bounded in a cuboid area, and the length, width and height of which are 30 meters, 30 meters and 6 meters, respectively. We set the initial user position as (-10, 0, 0). The user's trajectory is generated based on the conditional probability $	p(\mathbf{p}_{\mathrm{U}}^{(t)}|\mathbf{p}_{\mathrm{U}}^{(t-1)})=\mathcal{N}(\mathbf{p}_{\mathrm{U}}^{(t)};\mathbf{p}_{\mathrm{U}}^{(t-1)},\mathbf{C}_{\mathrm{p}})$, where $\mathbf{C}_{\mathrm{p}}$ is set as $\mathrm{diag}([0.03, 0.03, 0.01]^\mathrm{T})$. Therefore, the average movement distance of user between adjacent time slots is given by $\mathbb{E}\left[{\Vert\mathbf{p}_{\mathrm{U}}^{(t+1)}-\mathbf{p}_\mathrm{U}^{(t)}\Vert}_2\right] = \sqrt{0.07}$ m, which corresponds to a velocity of $48$ Km/h considering the 20ms time interval. The total number of tracking time slots is set as $T = 300$.\par
To evaluate the performance of the proposed BULT algorithm, we define the root mean square error (RMSE) of the estimated user position and AoAs respectively as
\begin{subequations}
	\label{62}
	\begin{align}
		&\mathrm{RMSE}\left(\mathbf{p}_\mathrm{U}\right)=\sqrt{\mathbb{E}\left[\frac{\sum_{t=1}^{T}{{\Vert\mathbf{p}_\mathrm{U}^{(t)}-\bar{\mathbf{p}}_\mathrm{U}^{(t)}\Vert}_2^2}}{T}\right]},\\
		&\mathrm{RMSE}\left({\theta}_{\mathrm{U},i}\right)=\sqrt{\mathbb{E}\left[\frac{\sum_{t=1}^{T}\sum_{i=1}^{K}{{\Vert{\theta}_{\mathrm{U},i}^{(t)}-\bar{{\theta}}_{\mathrm{U},i}^{(t)}\Vert}_2^2}}{TK}\right]}.
	\end{align}
\end{subequations}
The expectations in \eqref{62} are numerically approximated by averaging from 30 independently generated trajectories.
\begin{figure}[t]
	\centering
	\resizebox{8cm}{!}{\includegraphics[width=10cm]{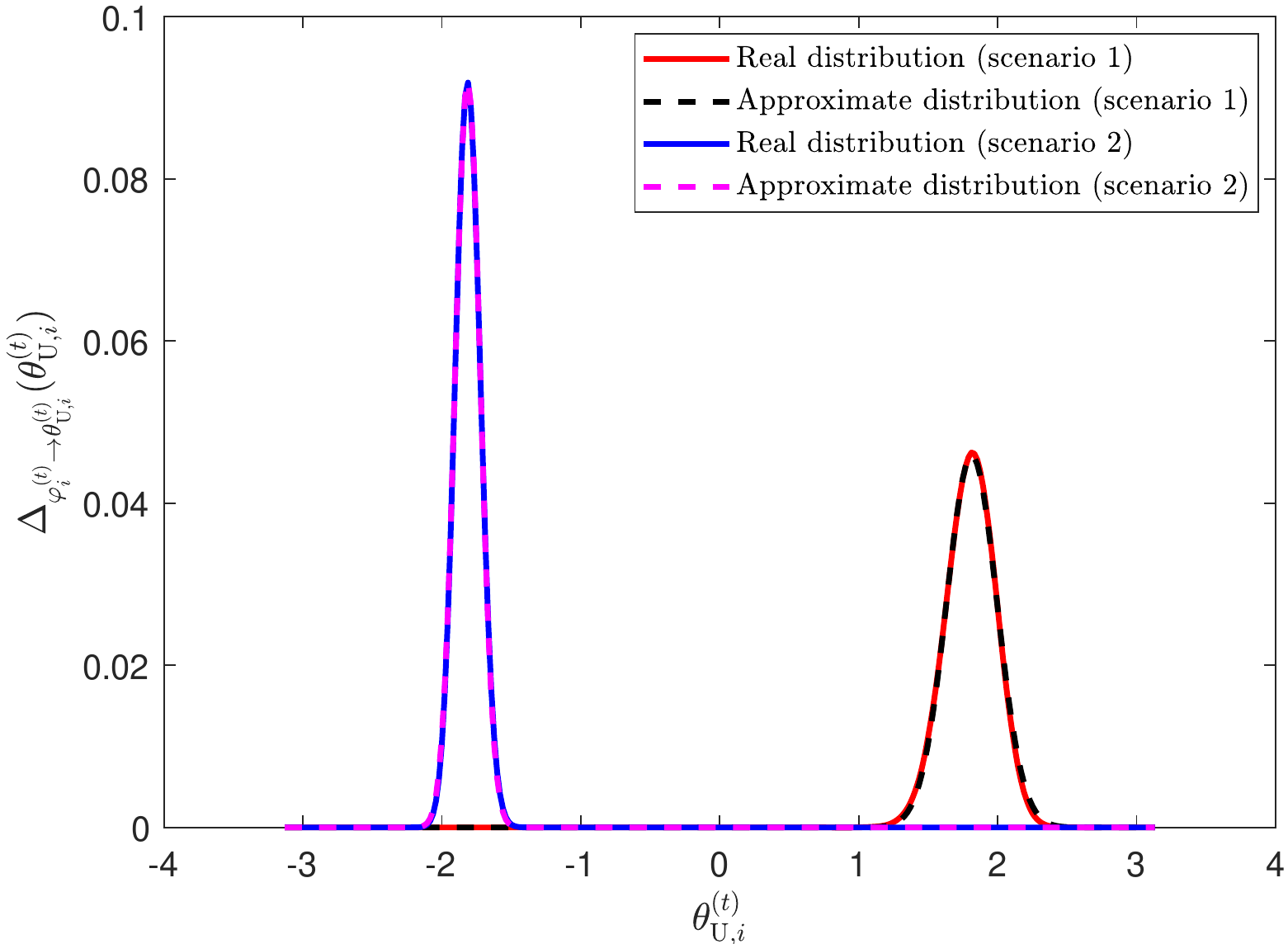}}
	\caption{Evaluation of the VM approximation in \eqref{VM_appro}, where the left peak and the right peak in the distribution correspond to scenario 1 and scenario 2 respectively.}
	\label{fig:4}
\end{figure}
\subsection{Message Approximation in \eqref{VM_appro}}
We conduct numerical experiments to compare the distribution given by \eqref{17} and its VM approximation given by \eqref{VM_appro}. The covariance matrix of Gaussian message $p(\theta_{\mathrm{U},i}^{(t)}|\mathbf{p}_{\mathrm{U}}^{(t)}) \Delta _{\mathbf{p}_{\mathrm{U}}^{(t)}\rightarrow\varphi_{i}^{(t)} }(\mathbf{p}_{\mathrm{U}}^{(t)})$ is set by $\mathbf{C}_{\mathbf{p}_{\mathrm{U}}^{(t)}\rightarrow\varphi_{i}^{(t)}} = \text{diag}({[2,2,2]^{\mathrm{T}}})$. We consider two scenarios: (1) the distance $d_i=40$ m and $\bar{\theta} _{\mathrm{U},i}^{(t)}=\frac{-\sqrt{3}\pi}{3}$; (2) the distance $d_i=20$ m and $\bar{\theta} _{\mathrm{U},i}^{(t)}=\frac{\sqrt{3}\pi}{3}$. As the results show in Fig. \ref{fig:4}, the approximation performs well when the far-field assumption is satisfied.
\begin{figure}[t]
	\centering
	\includegraphics[width=7cm]{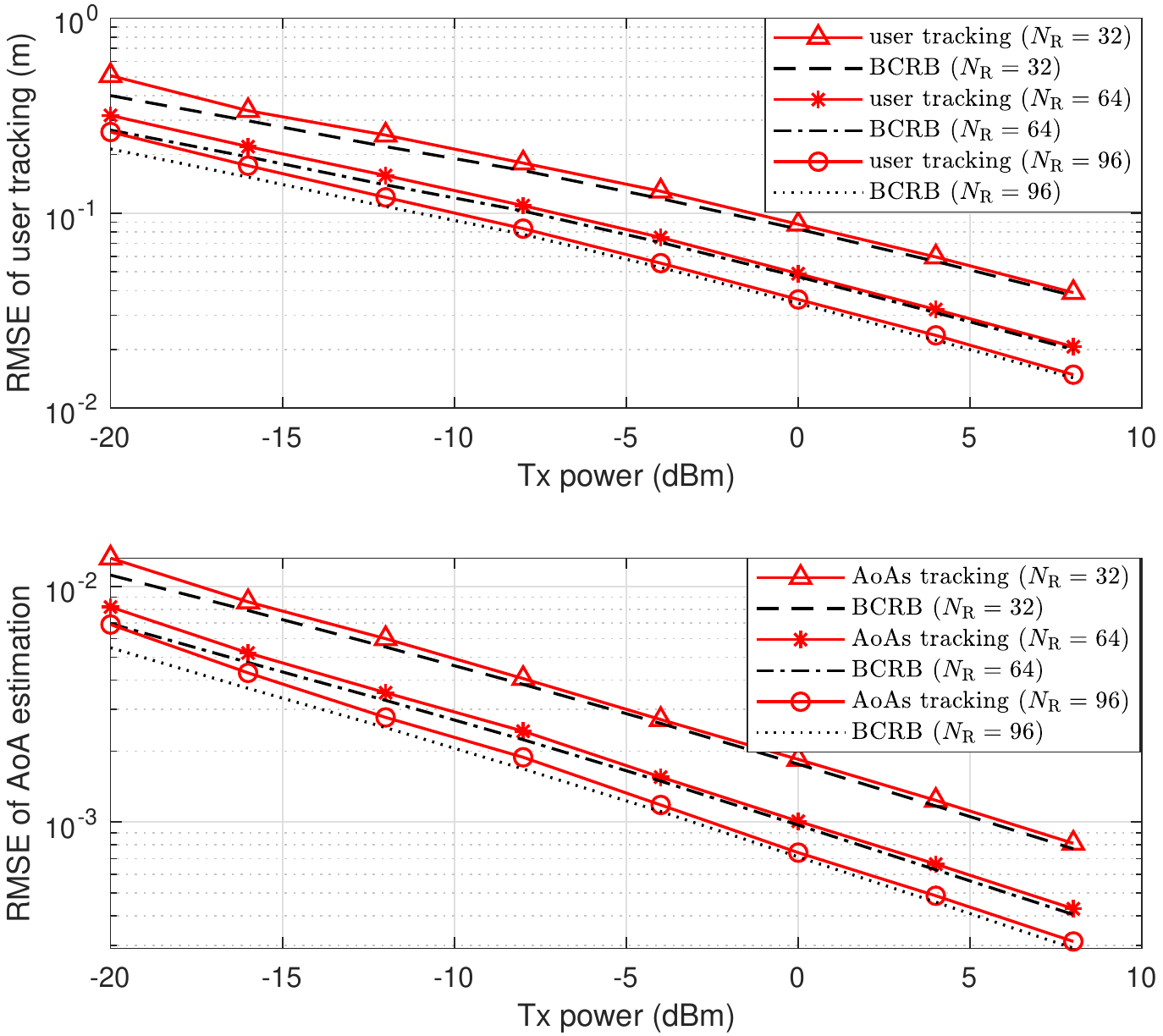}
	\caption{The user tracking and AoA estimation performance v.s. the transmission power with a varying number of RIS elements.}
	\label{fig:5}
\end{figure}
\subsection{Tracking and Estimation Performance}
In this subsection, we first evaluate the performance of the proposed BULT algorithm in different system parameter settings. We then compare the BULT algorithm with a straightforward tracking method \cite{2007Position} that acts as a baseline scheme. Finally, we perform experiments to compare the tracking performance under different beamforming designs given in Section \ref{Section5}.
\subsubsection{Tracking and AoAs Estimation}
\begin{figure}[t]
	\centering
	\includegraphics[width=7cm]{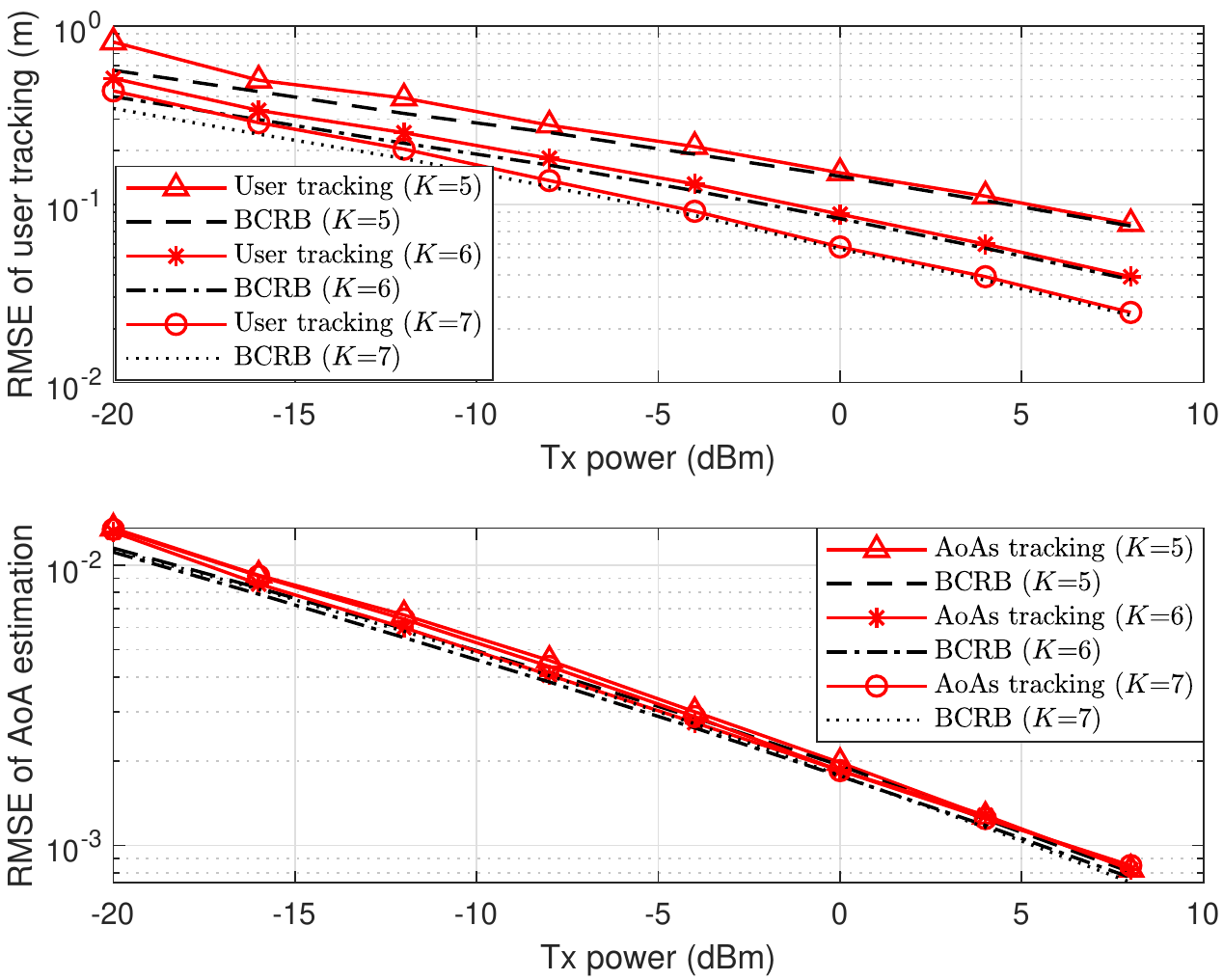}
	\caption{The user tracking and AoA estimation performance v.s. the transmission power with a varying number of RISs.}
	\label{fig:6}
\end{figure}

We study the impact of the number of RIS elements and the number of RISs on the user tracking performance in \eqref{62} with varying transmission power. The results are shown in Fig. \ref{fig:5} and Fig. \ref{fig:6}. For both the performance of user tracking and AoAs estimation, the increase of the RIS elements leads to performance improvement since the passive beamforming gain of the RISs is proportional to the number of the RIS elements. To gain insights into the relationship between the RIS number and the estimation error, we remove the RIS located in (-35, 5, -10) from the system with $K=6$ and remove the RISs located in (-35, 5, -10) and (10, 15, 20) for $K=5$. As given in Fig. \ref{fig:6}, the additional RISs in system brings better performance in user tracking while the AoAs estimation performances are almost the same. The additional RISs provide more information for user position which improves  the user tracking accuracy. However, it deteriorates the AoAs estimation performance slightly since more angle parameters are involved and need to be estimated. In Fig. \ref{fig:5} and Fig. \ref{fig:6}, it is observed that both the user tracking errors and the AoAs estimation errors are very close to the BCRB at relatively high SNR, which validates the effectiveness of the proposed BULT algorithm.
\subsubsection{Comparison with Benchmark}
\begin{figure}[t]
	\centering
	\includegraphics[width=7cm]{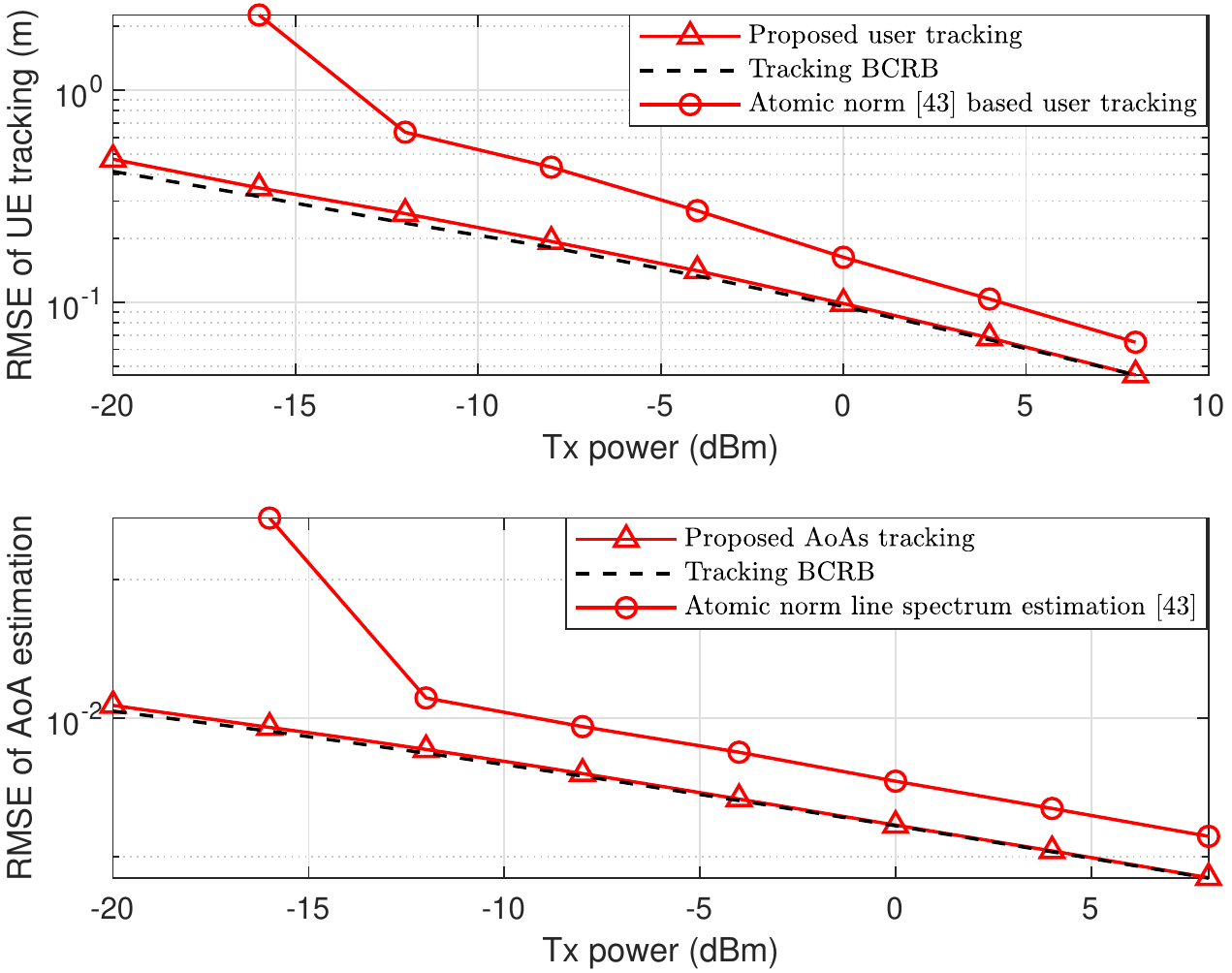}
	\caption{The user tracking and AoA estimation performances of the proposed BULT algorithm and its baseline schemes.}
	\label{fig:7}
\end{figure}

For the benchmark scheme, we consider the atomic norm based line spectrum inference method \cite{bhaskar2013atomic} to provide the estimation of AoAs. Then the user position is estimated successively based on the estimated AoAs by using the method in \cite{2007Position}. The results are given in Fig. \ref{fig:7}, where the proposed BULT algorithm significantly outperforms the atomic norm based user tracking. Our proposed algorithm considers the temporal correlation of the user position by building a probability model over time. The iterative message passing between the AoA estimation module and the position tracking module utilizes the geometric constraint in \eqref{eq10} which makes the estimation performance close to the BCRB.
\begin{figure}[t]
	\centering
	\includegraphics[width=7cm]{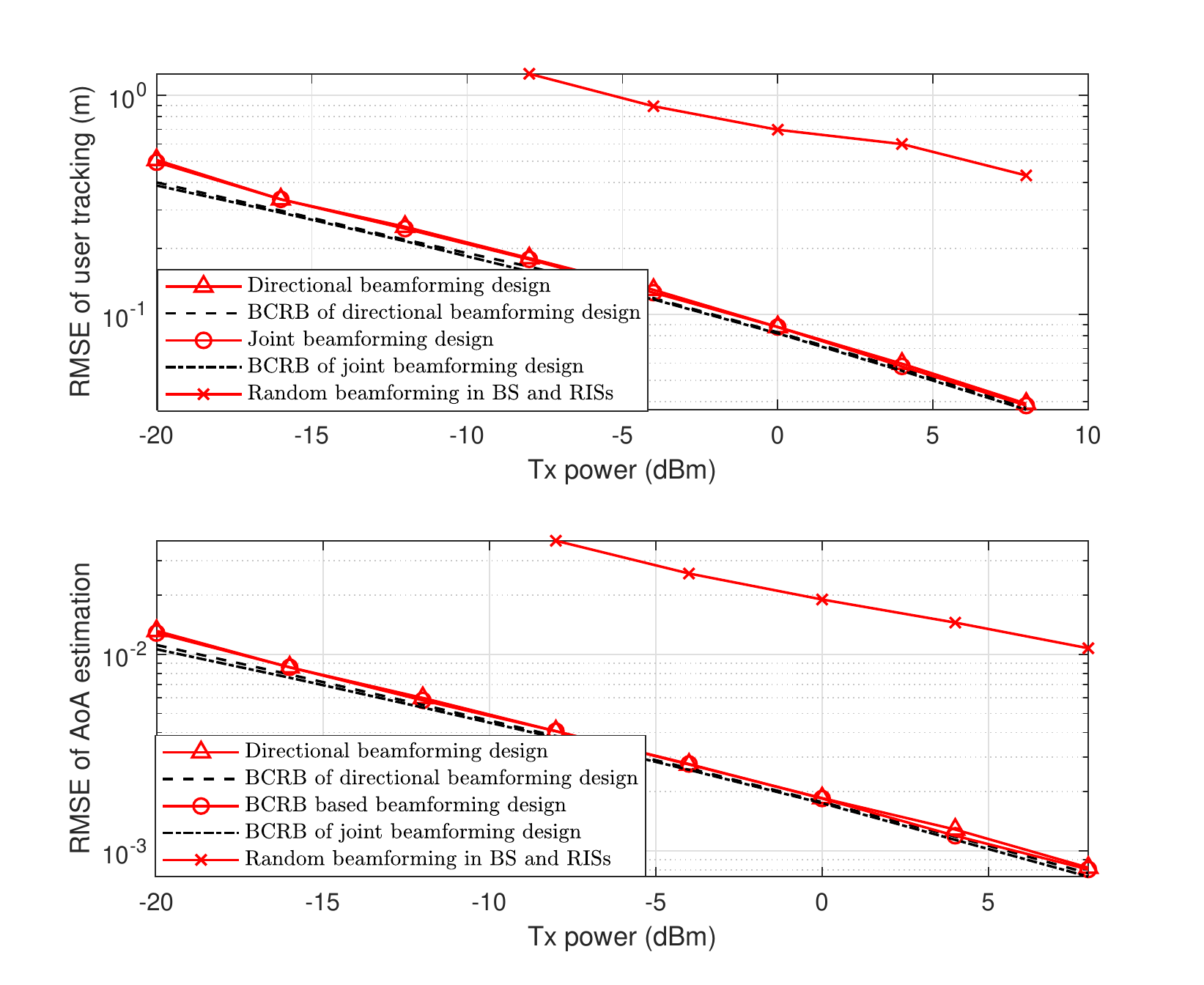}
	\caption{The user tracking and AoA estimation performance comparison of various beamforming methods}
	\label{fig:8}
\end{figure}
\subsubsection{Comparison of Beamforming Methods}
As shown in Fig. \ref{fig:8}, we compare the tracking performance of the directional beamforming design with that of the BCRB based beamforming design and the random beamforming design. The result shows that both the directional beamforming design and the BCRB based beamforming design significantly outperform the random beamforming design while the performance gap of these two superior beamforming designs is small for both user tracking and user AoA estimations. Therefore, it is preferable to use directional beamforming design in practice since it greatly reduces the computation complexity.
\section{Conclusion}
\label{Section7}
	In this paper, we studied the user localization and tracking problem in the RIS aided MIMO system. We utilized the geometric relationship between the user AoAs and the user position to establish the state transition model for the mobile user. Then, we proposed a novel message-passing based algorithm for online user tracking and AoAs estimation. Furthermore, we developed the BCRB to analyse the performance limits of the tracking problem which further guides the beamforming design of the BS and the RISs. Numerical simulation results show that the performance of the proposed BULT algorithm is close to the BCRB in user localization and tracking.

	
	%

	\appendices
	\section{Gaussian Message Approximation in \eqref{30_n}}
	\label{taylor_series}
	The expression of $\prod_{j=1}^{K}{\Delta_{\varphi_{i}^{(t)} \rightarrow \mathbf{p}_{\mathrm{U}}^{(t)}}(\mathbf{p}_{\mathrm{U}}^{(t)})}$ is given by
	\begin{align}
		\label{60}
		&\prod_{j=1}^{K}{\Delta_{\varphi_{i}^{(t)} \rightarrow \mathbf{p}_{\mathrm{U}}^{(t)}}(\mathbf{p}_{\mathrm{U}}^{(t)})}\notag\\
		&\quad\propto\exp\left(\sum_{j=1}^{K}{\kappa _{\theta_{\mathrm{U},j}^{(t)}\rightarrow \varphi_{j}^{(t)}}\cos(\pi(\mathbf{e}_{j}^{(t)})^{\mathrm{T}}\mathbf{e}_{\mathrm{U}}-\mu _{\theta_{\mathrm{U},j}^{(t)}\rightarrow \varphi_{j}^{(t)}})}\right),
	\end{align}
	where $\mathbf{e}_{j}^{(t)}$ is defined as $\mathbf{e}_{j}^{(t)}=\frac{\left( \mathbf{p}_{\mathrm{R},j}-\mathbf{p}_\mathrm{U}^{(t)} \right)}{\left\| \mathbf{p}_{\mathrm{R},j}-\mathbf{p}_{\mathrm{U}}^{(t)} \right\| _2}$. We resort to the gradient descent method to find the local maximum of \eqref{60} which is used as the mean vector $\mathbf{m}_{\mathcal{G}^{(t)}}$ of the approximated Gaussian message in \eqref{30_n}. The covariance matrix $\mathbf{C}_{\mathcal{G}^{(t)}}$ of the approximated Gaussian message is given by the Hessian matrix at $\mathbf{p}_{\mathrm{U}}^{(t)}=\mathbf{m}_{\mathcal{G}^{(t)}}$. We denote the exponential term of \eqref{60} as $f_e(\mathbf{p}_{\mathrm{U}}^{(t)})$. Thus, the gradient of $f_e(\mathbf{p}_{\mathrm{U}}^{(t)})$ is derived in
	\begin{align}
		\frac{\partial{ f_e(\mathbf{p}_{\mathrm{U}}^{(t)})}}{\partial\mathbf{p}_{\mathrm{U}}^{(t)}}\!=\!\sum_{j=1}^{K}\!{\left(-\pi\kappa \!_{\theta_{\mathrm{U},j}^{(t)}\rightarrow \varphi_{j}^{(t)}}\sin\!\left(\!\pi(\mathbf{e}_{j}^{(t)})^{\mathrm{T}}\mathbf{e}_{\mathrm{U}} \!-\!\mu\! _{\theta_{\mathrm{U},j}^{(t)}\rightarrow \varphi_{j}^{(t)}}\!\right)\!\mathbf{u}_j^{(t)}\!\right)},
	\end{align}
with $\mathbf{u}_j^{(t)}=\frac{-\mathbf{e}_\mathrm{U}+({\mathbf{e}_{j}^{(t)}})^{T}\mathbf{e}_\mathrm{U}\mathbf{e}_{j}^{(t)}}{\left\| \mathbf{p}_{\mathrm{R},j}-\mathbf{p}_{\mathrm{U}}^{(t)} \right\| _2}$. We iteratively calculate
\begin{equation}
	\label{64}
	\mathbf{p}_{\mathrm{U},(i+1)}^{(t)} = \mathbf{p}_{\mathrm{U},(i)}^{(t)} + \lambda\left. \frac{\partial f_e(\mathbf{p}_{\mathrm{U}}^{(t)})}{\partial \mathbf{p}_{\mathrm{U}}^{(t)}} \right|_{\mathbf{p}_{\mathrm{U}}^{(t)}=\mathbf{p}_{\mathrm{U},(i)}^{(t)}},
\end{equation}
where $\lambda$ is a predetermined step size, and $i$ is the iteration index. The iteration stops when it satisfies the stopping criterion. $\mathbf{m}_{\mathcal{G}^{(t)}}$ is set by the obtained local optimum.
For the calculation of $\mathbf{C}_{\mathcal{G}^{(t)}}$, the Hessian matrix of $f_e(\mathbf{p}_{\mathrm{U}}^{(t)})$ is expressed as
\begin{align}
	&\frac{\partial^2f_e(\mathbf{p}_{\mathrm{U}}^{(t)})}{\partial\mathbf{p}_{\mathrm{U}}^{(t)}\partial(\mathbf{p}_{\mathrm{U}}^{(t)})^\mathrm{T}} \notag\\
	&=\!\sum_{j=1}^{K}{\Biggl(\!-\pi^2\kappa _{\theta_{\mathrm{U},j}^{(t)}\rightarrow \varphi_{j}^{(t)}}\cos\left(\pi(\mathbf{e}_{j}^{(t)})^{\mathrm{T}}\mathbf{e}_{\mathrm{U}}\!-\!\mu _{\theta_{\mathrm{U},j}^{(t)}\rightarrow \varphi_{j}^{(t)}}\right)\mathbf{u}_j^{(t)}(\mathbf{u}_j^{(t)})^{\mathrm{T}}\!}\notag\\
	&\quad-\pi\kappa _{\theta_{\mathrm{U},j}^{(t)}\rightarrow \varphi_{j}^{(t)}}\sin\left(\pi(\mathbf{e}_{j}^{(t)})^{\mathrm{T}}\mathbf{e}_{\mathrm{U}}\!-\!\mu _{\theta_{\mathrm{U},j}^{(t)}\rightarrow \varphi_{j}^{(t)}}\right)\!\notag\\
	&\quad{\left.\times\left(\!\frac{3\mathbf{e}_\mathrm{U}^\mathrm{T}\mathbf{e}_{j}^{(t)}\mathbf{e}_{j}^{(t)}{\mathbf{e}_{j}^{(t)}}^\mathrm{T}}{\left\| \mathbf{p}_{\mathrm{R},j}\!-\!\mathbf{p} \right\| _2^2}\!-\!\frac{\mathbf{e}_\mathrm{U}^\mathrm{T}\mathbf{e}_{j}^{(t)}}{\left\| \mathbf{p}_{\mathrm{R},j}\!-\!\mathbf{p} \right\| _2^2}\mathbf{I}\!\right)\!\right)}.
\end{align}
Then, we approximate $\mathbf{C}_{\mathcal{G}^{(t)}}$ as
\begin{equation}
	\mathbf{C}_{\mathcal{G}^{(t)}} = \left(-\left. \frac{\partial ^2f_e(\mathbf{p}_{\mathrm{U}}^{(t)})}{\partial \mathbf{p}_{\mathrm{U}}^{(t)}\partial (\mathbf{p}_{\mathrm{U}}^{(t)})^{\mathrm{T}}} \right|_{\mathbf{p}_{\mathrm{U}}^{(t)}=\mathbf{m}_{\mathcal{G}^{(t)}}}
\right)^{-1}.
\end{equation} 
\section{Derivation of \eqref{VM_appro}}
	\label{VMapproximation_derivation}
	For notational brevity, we denote $\mathbf{m}_{\mathbf{p}_{\mathrm{U}}^{(t)}\rightarrow\varphi_{i}^{(t)}}$ and $\mathbf{C}_{\mathbf{p}_{\mathrm{U}}^{(t)}\rightarrow\varphi_{i}^{(t)}}$ by $\hat{\mathbf{p}}_{\mathrm{U},i}^{(t)}$ and $\hat{\mathbf{C}}_{\mathrm{U},i}^{(t)}$ respectively. We simplify the integral of \eqref{17} by projecting the user position error $(\mathbf{p}_{\mathrm{U}}^{(t)}-\hat{\mathbf{p}}_{\mathrm{U},i}^{(t)})$ on the directional vector $\mathbf{v}_i$. We represent the projection of $(\mathbf{p}_{\mathrm{U}}^{(t)}-\hat{\mathbf{p}}_{\mathrm{U},i}^{(t)})$ onto $\mathbf{v}_i$ by a random variable $x_i$ following the distribution of $\mathcal{N}(x_i;0,\mathbf{v}_i^\mathrm{T}\hat{\mathbf{C}}_{\mathrm{U},i}^{(t)}\mathbf{v}_i)$. Recall from Fig. \ref{fig_VM_approximate} that $\mathbf{v}_i$ is perpendicular to $\mathbf{p}_{\mathrm{R},i}-\mathbf{m}_{\mathbf{p}_{\mathrm{U}}^{(t)}\rightarrow\varphi_{i}^{(t)}}$. Thus, for a sufficiently large $d_i =\left\|\mathbf{p}_{\mathrm{R},i}-\mathbf{m}_{\mathbf{p}_{\mathrm{U}}^{(t)}\rightarrow \theta _{\mathrm{U},i}^{(t)}\mathbf{p}_{\mathrm{U}}^{(t)}}\right\|_2$, the geometric constraint in \eqref{eq10} reduces to
	\begin{equation}
		\tan(\arccos{\bar{\theta}_{\mathrm{U},i}^{(t)}}-\arccos{\theta _{\mathrm{U},i}^{(t)}})=\frac{x_i}{d_i},
	\end{equation}
	 Hence, the message in \eqref{17} is approximated as
	 \begin{subequations}
	 \begin{align}
	 	&\Delta _{\varphi_{i}^{(t)}\rightarrow \theta_{\mathrm{U},i}^{(t)}}(\theta_{\mathrm{U},i}^{(t)})\notag\\
	 	&\quad\!\propto\!\int_{x_{i}}\!{\delta\!\left(\frac{x_i}{d_i}\!-\!\tan(\arccos{\bar{\theta} _{\mathrm{U},i}^{(t)}}\!-\!\arccos{\theta _{\mathrm{U},i}^{(t)}})\right)\!\exp\!\left(\!-\frac{x_i^{2}}{2\mathbf{v}_i^\mathrm{T}\!\hat{\mathbf{C}}_{\mathrm{U},i}^{(t)}\!\mathbf{v}_i}\!\right)}\\
	 	&\quad\propto\exp(-\frac{d_i^2\tan^2(\arccos{\bar{\theta}} _{\mathrm{U},i}^{(t)}-\arccos{\theta _{\mathrm{U},i}^{(t)}})}{2\mathbf{v}_i^\mathrm{T}\hat{\mathbf{C}}_{\mathrm{U},i}^{(t)}\mathbf{v}_i}).\label{33}
	 \end{align}
	 \end{subequations}
	\par 
	The message in \eqref{33} achieves its maximum at $\theta_{\mathrm{U},i}^{(t)} = \bar{\theta} _{\mathrm{U},i}^{(t)}$. Since the VM distribution in \eqref{VM_appro} is maximized when $\pi\theta_{\mathrm{U},i}^{(t)} = \mu _{\varphi_{i}^{(t)}\rightarrow \theta_{\mathrm{U},i}^{(t)}}$, we set $\mu _{\varphi_{i}^{(t)}\rightarrow \theta_{\mathrm{U},i}^{(t)}} = \pi\bar{\theta} _{\mathrm{U},i}^{(t)}$. As for $\kappa _{\varphi_{i}^{(t)}\rightarrow \theta_{\mathrm{U},i}^{(t)}}$, we take the Taylor series expansion of \eqref{VM_appro} and \eqref{33} at the maximum with its second-order derivative given by
	\begin{align}
			\label{talorexpon1}
			&\frac{\mathrm{d}^2\log \left( \Delta _{\varphi_{i}^{(t)}\rightarrow \theta_{\mathrm{U},i}^{(t)}}(\theta _{\mathrm{U},i}^{(t)}) \right)}{\mathrm{d}{\theta _{\mathrm{U},i}^{(t)}}^2}\notag\\
			&\!=\!\frac{-d_{i}^{2}}{\mathbf{v}_{i}^{\mathrm{T}}\!\mathbf{\hat{C}}_{\mathrm{U},i}^{(t)}\!\mathbf{v}_i\sin ^2\!\left( \xi \right)}\!\left(\! \frac{1 \!+\!2\sin ^2\!\left( \!\bar{\xi}_{i}^{(t)}\!-\!\xi_{i}^{(t)}\! \right)}{\cos ^4\left(\! \bar{\xi}_{i}^{(t)}\!-\!\xi_{i}^{(t)} \!\right)}\!-\!\frac{\sin \left(\! \bar{\xi}_{i}^{(t)}\!-\!\xi_{i}^{(t)} \!\right)\!\cos\!\xi}{\cos ^3\!\left(\!\bar{\xi}_{i}^{(t)}\!-\!\xi_{i}^{(t)}\!\right)\!\sin\! \xi} \!\right),\\
			\label{talorexpon2}	
			&\frac{\mathrm{d}^2\log \left( \mathcal{M}\left(\pi\theta_{\mathrm{U},i}^{{(t)}};\mu_{\varphi_{i}^{(t)}\rightarrow \theta_{\mathrm{U},i}^{(t)}},\kappa_{\varphi_{i}^{(t)}\rightarrow \theta_{\mathrm{U},i}^{(t)}}\right) \right)}{\mathrm{d}{\theta _{\mathrm{U},i}^{{(t)}}}^{2}}\notag\\
			&\quad=-\pi^2\kappa_{\varphi_{i}^{(t)}\rightarrow \theta_{\mathrm{U},i}^{(t)}}\cos(\pi\theta _{\mathrm{U},i}^{(t)}-\mu _{\varphi_{i}^{(t)}\rightarrow \theta_{\mathrm{U},i}^{(t)}}),
	\end{align}
	where $\xi_{i}^{(t)} = \arccos(\theta_{\mathrm{U},i}^{(t)})$ and $\bar{\xi}_{i}^{(t)} = \arccos(\bar{\theta}_{\mathrm{U},i}^{(t)})$. By letting 
	\begin{flalign}
		&\left.\frac{\mathrm{d}^2 \log \left( \Delta  _{\varphi _{i}^{( t)}\rightarrow\theta _{\mathrm{U},i}^{(t)}}(\theta _{\mathrm{U},i}^{(t)})\right)}{\mathrm{d}{\theta _{\mathrm{U},i}^{(t)}}^2}\right|_{\theta _{\mathrm{U},i}^{(t)}=\bar{\theta}_{\mathrm{U},i}^{(t)}}\notag\\
		& \quad =\left.\frac{\mathrm{d}^2\log \left(\mathcal{M}\left(\pi\theta _{\mathrm{U},i}^{(t)};\mu_{\varphi _{i}^{(t)}\rightarrow \theta _{\mathrm{U},i}^{(t)}},\kappa_{\varphi _{i}^{(t)}\rightarrow \theta _{\mathrm{U},i}^{(t)}} \right) \right)}{\mathrm{d}{\theta _{\mathrm{U},i}^{(t)}}^2}\right|_{\theta _{\mathrm{U},i}^{(t)}=\bar{\theta}_{\mathrm{U},i}^{(t)}},
	\end{flalign}
	we obtain $\kappa_{\varphi_{i}^{(t)}\rightarrow \theta_{\mathrm{U},i}^{(t)}}=\frac{d_i^2}{\pi^2\sin^2{\bar{\xi}_{i}^{(t)}}\mathbf{v}_i^\mathrm{T}\hat{\mathbf{C}}_{\mathrm{U},i}^{(t)}\mathbf{v}_i}$.
	\ifCLASSOPTIONcaptionsoff
	\newpage
	\fi

	
	
	\bibliographystyle{IEEEtran}
	\bibliography{newbib}
\end{document}